\documentclass[12pt]{iopart}

\usepackage{graphicx}
\usepackage{cite}
\usepackage{hyperref}
\usepackage{etoolbox}

\makeatletter
\def\@mkboth#1#2{}
\newlength\appendixwidth
\preto\appendix{\addtocontents{toc}{\protect\patchl@section}}
\newcommand{\patchl@section}{%
  \settowidth{\appendixwidth}{\textbf{Appendix }}%
  \addtolength{\appendixwidth}{1.5em}%
  \patchcmd{\l@section}{1.5em}{\appendixwidth}{}{\ddt}%
}
\makeatother

\begin{document}
\bibliographystyle{iopart-num}

\begin{flushright}
	SAGEX-22-13\\
\end{flushright}

\title[Amplitudes and collider physics]{The SAGEX Review on Scattering Amplitudes \\ Chapter 12: Amplitudes and collider physics}

\author{Chris D. White}

\address{Centre for Theoretical Physics, Department of Physics and
  Astronomy,\\ Queen Mary University of London, Mile End Road, London
  E1 4NS,\\ United Kingdom} \ead{christopher.white@qmul.ac.uk}
\vspace{10pt}
\begin{indented}
\item[]October 2021
\end{indented}

\begin{abstract}
We explore how various topics in modern scattering amplitudes research
find application in the description of collider physics
processes. After a brief review of experimentally measured quantities
and how they are related to amplitudes, we summarise recent
developments in perturbative QFT, and how they have impacted our
ability to do precision physics with colliders. Next, we explain how
the study of (next-to-)soft radiation is directly relevant to
increasing theoretical precision for key processes at the LHC and
related experiments. Finally, we describe the various techniques that
are used to turn theoretical calculations into something more closely
approaching the output of a particle accelerator. 
\end{abstract}

%
%
%
%
%

\tableofcontents

\section{Introduction}
\label{sec:intro}

As the SAGEX network has made clear, the study of scattering
amplitudes has become an extraordinarily broad endeavour. Theories
studied range from well-established theories of nature (e.g. the
Standard Model, plus effective field theory corrections) right through
to highly supersymmetric versions of non-abelian gauge theories, or
even string theories. The latter are less relevant for current
experiments, but can be useful for developing new calculational
techniques, or for showing up relationships between different types of
theory. Furthermore, new mathematics is often needed in order to bring
new aspects of field theory to light, such that dialogue between pure
mathematicians and physicists has been an increasing facet of the
amplitudes community in recent years. What makes this mix so important
is that new techniques used to study more formal theories are very
often {\it the same techniques} that have been used to generate highly
non-trivial results in more physically relevant theories. However, the
diversity of the amplitudes community can be as much of a hindrance as
it is a strength: it becomes ever-more difficult for any given person
-- and especially a newcomer to the field -- to see how different
branches of amplitudes research are related, and to know how formal
ideas in one corner are applied to very practical ends in another.

The aim of this article is to redress this balance somewhat, by
focusing on one of the main motivations for studying amplitudes in the
first place, namely that they underlie the physics of collider
experiments. The latter form our main way of testing theories of
fundamental physics, which itself is at a crucial crossroads. With the
discovery of the Higgs boson in 2012, the SM is experimentally
complete, but its obvious deficiencies\footnote{Examples include the
observed matter / antimatter asymmetry in our universe, quantum
instability of the Higgs boson mass (the hierarchy problem), the
absence of gravity, lack of explanation of particle masses, dark
matter or dark energy, and much else besides.} cry out for a deeper
explanation. The current flagship experiment is the Large Hadron
Collider at CERN, which will be with us for at least another two
decades. Replacement colliders are being actively considered, although
the lack of any particularly striking new physics hints at the LHC
make it rather unclear what the optimum follow-up machine will look
like.

In order to make progress, we must continue to develop new types of
collider, as well as our understanding of quantum field theory. On
very general grounds, any new physics will look QFT-like at currently
available collider energies. Furthermore, the lack of any very clear
new physics means that we have to understand the old physics (i.e. the
SM) extremely well. Only by knowing the old physics {\it very very
  precisely} can we be sure that tiny deviations we might see in
experiments are genuine new physics, rather than poorly understood old
physics. Furthermore, given that collider experiments involve
scattering particles, it is the study of {\it scattering amplitudes}
in QFT that is the most relevant thing to be doing.

Given that you are reading this article, I am assuming that you will
already be familiar with what a scattering amplitude is, whether or
not this is in our own apparently four-dimensional and distinctly
non-conformal world. However, you may well not be familiar with what
goes on inside a collider, what experimentalists do with it, and how
you can start to turn a scattering amplitude into something they might
be interested in. Although some of the ideas may well have already
been presented to you in some region of your past light-cone, we
review relevant material in the following sections.

\section{What is a collider, and what does it measure?}
\label{sec:collider}

A {\it particle accelerator} or {\it collider} is an experiment that
accelerates one or more beams of particles so that they collide. Early
versions of this idea had a single beam colliding with a fixed target,
which is easier to build, but suffers from the fact that much of the
energy in the initial state gets wasted as kinetic energy in the final
state. Hence, for the past few decades, colliders have consisted of
two beams of particles that collide in the centre-of-mass
frame. Electric and magnetic fields are used to accelerate the
particles and focus the beams, and thus the particles being
accelerated must be (electromagnetically) charged. Recent examples
include LEP ($e^+ e^-$), the Tevatron ($p\bar{p}$), and the LHC
($pp$), where $e^\pm$ denotes the electron and positron, and
$(p,\bar{p}$) the (anti-)proton. Colliders tend to be circular, which
gives multiple chances to accelerate the particles each time they go
around the ring. However, charged particles of energy $E$ and mass $m$
radiate {\it synchrotron radiation} at a rate $\propto
(E/m)^{4}r^{-2}$, imposing a practical limit on how much we can
accelerate them. This is why the most modern collider (the LHC) uses
heavy particles (protons rather than electrons), and has a very large
circumference of 27 km!  Protons are not ideal -- they are composite,
wobbly bags of quarks and gluons. This makes hadron colliders messier
than $e^- e^+$ colliders, making precision physics more
difficult. Thus, future colliders may return to using electrons and
positrons, at the expense of needing new technology for colliding them
in a {\it linear} fashion.

The beams in a particle accelerator are focused so that they collide,
which in modern circular colliders happens at multiple points around
the ring. Each collision of the beam particles is referred to as a
{\it (scattering) event}, and the aim of each experiment is to record
and analyse interesting-looking events. The collision points are
surrounded by a cylindrical detector, whose aim is to collect as many
of the particles emerging from a collision point as possible.  The
detectors contain multiple layers for capturing different types of
particle, and measuring their momenta and charges. Some particles,
such as weakly-interacting neutrinos or possible new physics
particles, pass through the detector and thus appear as ``missing
4-momentum'' (i.e. a failure of momentum conservation) in a given
scattering event. From a very na\"{i}ve theorist's viewpoint, one may
thus characterise individual scattering events by a recorded list of
particles, together with their measured 4-momenta. We will return to
how accurate this picture is in section~\ref{sec:experiment}, but for
now let us note that in a quantum theory it is not possible to predict
exactly which particles will emerge in any given event, nor what their
precise momenta will be. Instead, we can only calculate {\it
  probabilities} for certain final states to appear in the detector,
and then compare the measured properties of events with predicted
distributions from a given QFT.

If we consider a simple classical scattering process, such as throwing
tennis balls at a target, it is clear that the probability of hitting
the target depends on its cross-sectional area. This is equally true
of quantum scattering, for which the relevant cross-sectional area
represents an effective range of interaction for particles incident on
the target. It follows that the rate of scattering events for a given
pair of incoming beams can be written as 
\begin{equation}
\frac{dN}{dt}\propto \sigma,
\label{dNdt}
\end{equation}
where $N$ is the number of events, $t$ the time, and $\sigma$ has
units of area. It is called the {\it cross-section}, and depends on
the intrinsic properties of the incoming particles, thus is calculable
from QFT. The quantity relating the two sides of eq.~(\ref{dNdt}) is
usually written ${\cal L}(t)$, as it may depend on time in general. It
is called the {\it luminosity} and, roughly speaking, measures how the
incoming particles are distributed within the beams i.e. how
``bright'' the beams are. Here we have been talking about the total
number of scattering events, but we can choose to focus on a
particular type of event e.g. those containing a pair of top quarks,
or a single Higgs boson. Letting $N_i$ denote the number of events for
distinct processes $\{i\}$, we have
\begin{equation}
N=\sum_i N_i\quad\Rightarrow\quad \frac{dN_i}{dt}={\cal L}(t)\sigma_i,
\label{Nidef}
\end{equation}
where $\sigma_i$ is the cross-section for process $i$, such that the
total cross-section for interaction is $\sigma=\sum_i \sigma_i$. It is
not possible to think of each individual cross-section as a physical
area, but we do not have to do this: eq.~(\ref{Nidef}) tells us that
the cross-section for a given process is simply given by the event
rate, divided by the luminosity. Experimentalists measure the
luminosity ${\cal L}(t)$, so that they can then present results for
cross-sections directly, for comparison with theory. For historical
reasons, the conventional unit of cross-section is the {\it barn} (b),
where one barn is defined to be 10$^{-28}$m$^2$. For context, the total $pp$
cross-section at the LHC (at 13 TeV) is about 0.1b. The cross-section
for single Higgs boson production in its main production mode
(gluon-gluon fusion) is about 57pb (here the ``p'' stands for {\it
  pico}, or $10^{-12}$), which gives some indication of the huge range
of cross-sections for interesting processes at the LHC, and how very
tiny signals have to be extracted from a massive amount of other
data. Given that the luminosity is related to the total event rate, it
follows that the {\it integrated luminosity}
\begin{displaymath}
  L(t)=\int_0^t dt' {\cal L}(t')
  \label{Lint}
\end{displaymath}
is a measure of the total number of events collected by a given
collider after a time of operation $t$. From eq.~(\ref{dNdt}) and the
fact that the event number $N$ is dimensionless, the integrated
luminosity has units of inverse area. As an example, the LHC delivered
156 fb$^{-1}$ (inverse femtobarns) of integrated luminosity during Run
2 at 13 TeV. Given the inverse units, a smaller prefix (nano, pico,
femto) means more events have been collected.

\section{Cross-sections and amplitudes}
\label{sec:amplitudes}

In the previous section, we have seen that a given scattering process
$i$ has an associated cross-section $\sigma_i$. In principle, QFT
tells us how to calculate this. Let us first consider the case that
the incoming and outgoing particles are fundamental particles. Let
$p_1$ and $p_2$ be the incoming particle momenta, and $\{p_3,\ldots
p_n\}$ the final momenta, so that there are $(n-2)$ particles in the
final state. Then your QFT textbook will tell you that the
cross-section can be decomposed into the following form:
\begin{equation}
  \sigma_i=\frac{1}{F(p_1,p_2)}\int
  d\Phi^{(n-2)}(p_3,\ldots,p_n) {\cal A}_i(\{p_i\}).
\label{cross-sec}
\end{equation}
Here
\begin{equation}
  F(p_1,p_2)=4[(p_1\cdot p_2)^2-m_1^2m_2^2]^{1/2},
  \label{flux}
\end{equation}
where $m_i$ is the mass of beam particle $i$, is called the {\it
  Lorentz-invariant flux}. It is a convenient relativistic
generalisation of the usual flux describing how the beam particles are
moving relative to each other. There is also an integration over the
momenta of the final state particles, given by the {\it
  Lorentz-invariant phase space}
\begin{equation}
  \int d\Phi^{(n-2)}= (2\pi)^4 \left(\prod_{i=1}^{n-2}
  \int \frac{d^3 \vec{p}_i}{(2\pi)^3 2E_i}\right)
  \delta^{(4)}(P_{\rm final}-P_{\rm init.}).
  \label{dPhidef}
\end{equation}
Here $E_i$ is the energy of particle $i$, and $\vec{p}_i$ its
(relativistic) 3-momentum. Also, $P_{\rm init.}$ ($P_{\rm final}$) are
the total 4-momentum in the initial (final) state respectively. We
then see that the delta function in eq.~(\ref{dPhidef}) implements
total momentum conservation in the scattering process. Finally in
eq.~(\ref{cross-sec}), ${\cal A}_i$ is the {\it scattering amplitude}
for process $i$, which the SAGEX collaboration was set up to
investigate in a myriad of ways. Equation~(\ref{cross-sec}) suggests
how you as a theorist might compare a given QFT with experiment: for a
given process, you can calculate the appropriate scattering amplitude
using perturbation theory, before integrating over the phase-space,
and dividing by the appropriate flux factor. The resulting number may
then be compared with something produced by experimentalists, if they
have isolated the relevant events. The process might be a potential
new physics process, and if the number agrees you may end up with an
expenses-paid trip to Stockholm. Or it might be a process that occurs
in the Standard Model, so that your number allows an experimentalist
to efficiently disentangle new from old physics, so that they might go
to Stockholm.

\begin{figure}
  \begin{center}
    \scalebox{0.6}{\includegraphics{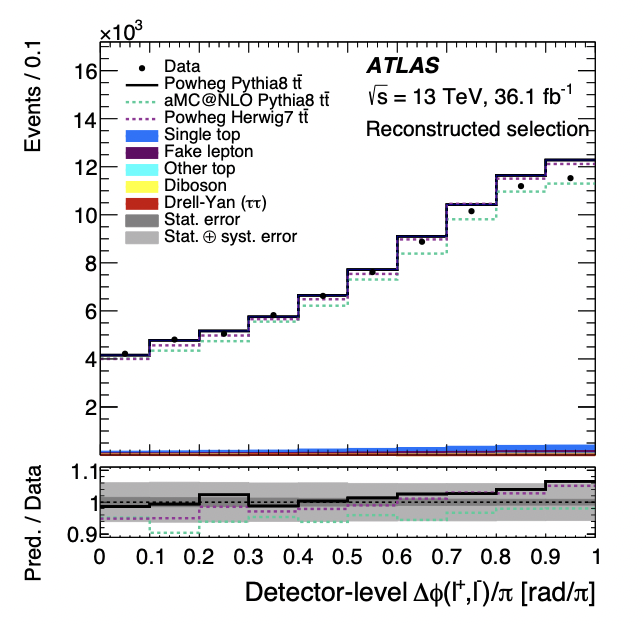}}
    \caption{Distribution of the azimuthal angle between leptons
      originating from top quark decays. From
      ref.~\cite{ATLAS:2019zrq}.}
    \label{fig:phiplot}
  \end{center}
\end{figure}

Total cross-sections are a very crude way of comparing
theory with data. It is also possible to compare {\it differential
  cross-sections}, where one does not integrate over the full phase
space in eq.~(\ref{cross-sec}). Given an observable ${\cal O}$, the
quantity $(d\sigma_i/d{\cal O})$ represents the distribution of ${\cal
  O}$ as measured in events collected in scattering process
$i$. Experimentalists often normalise this by dividing by the total
cross-section, given that many uncertainties cancel out when doing
this. An example distribution is shown in figure~\ref{fig:phiplot},
which comes from an analysis of events containing (anti-)top quark
pairs by the ATLAS experiment~\cite{ATLAS:2019zrq}. The (anti-)top
quarks decay to produce leptons, and the collaboration have measured
the azimuthal angle in the detector between these decay products,
where the data is compared to various predictions from
theorists. Interestingly, the theory curves do not agree with the data
particularly well by eye. Statistically, however, there is no cause
for excitement: the lower panel depicts the uncertainty on the
measurement as a grey band, which indeed overlaps with the (Standard
Model) theory.

It goes almost without saying that the above discussion is vastly
oversimplified. The comparison of theory with data is, in practice,
almost nothing like the simple procedure outlined here. Typical
complications include the following:
\begin{itemize}
\item Amplitudes in QFT suffer from ultraviolet (UV) divergences at
  loop-level. These can be dealt with via renormalisation in some
  scheme, and modern-day theorists typically use {\it dimensional
    regularisation} in $4-2\epsilon$ dimensions. Then cross-sections
  for comparison with experiment will depend on the renormalisation
  scale $\mu_R$. In particular, the QCD expansion parameter will
  become a {\it running coupling} $\alpha_s(\mu_R)$. The coefficients
  of the perturbation expansion in this parameter involve large
  logarithms involving ratios of $\mu_R$ with energy scales typical of
  the scattering process (e.g. invariants formed from the particle
  momenta). Thus, dependence on $\mu_R$ is minimised at a given order
  in perturbation theory by choosing $\mu_R$ to be comparable with
  these energy scales.
\item The incoming particles at the LHC are protons, which are not
  fundamental particles in QCD, and thus not fully described by
  standard perturbation theory. We must then generalise the formula of
  eq.~(\ref{cross-sec}), given that we can calculate scattering
  amplitudes in perturbation theory for incoming (anti-)quarks and /
  or gluons, but not protons. We discuss this in
  section~\ref{sec:partons}.
\item We can typically only calculate the amplitude to low orders in
  perturbation theory, where the strong interaction (with expansion
  parameter $\alpha_s\equiv g_s^2/(4\pi)$, where $g_s$ is the QCD
  coupling constant) is the most relevant. The state of the art for
  many processes of interest is next-to-leading order (NLO) in
  $\alpha_s$, although NNLO is known for some cases. Some
  cross-sections are known to a highly impressive N$^3$LO! Examples
  include different production modes for the Higgs
  boson~\cite{Anastasiou:2015vya,Anastasiou:2016cez,Dreyer:2016oyx,Dreyer:2018qbw},
  and the so-called {\it Drell-Yan} production of a heavy
  particle~\cite{Duhr:2020seh}, which we will see in more detail
  below.  Proceeding order-by-order in the coupling in this way is
  known as {\it fixed-order perturbation theory}, and we discuss this
  in section~\ref{sec:fixedorder}.
\item Perturbation theory can be unstable: for differential
  cross-sections, the coefficients of the perturbation series depend
  on the external particle momenta, and can diverge in certain
  kinematic regions (such as the {\it soft limit}, in which the
  4-momentum of emitted radiation goes to zero). One must then sum up
  certain contributions to all orders in perturbation theory for
  meaningful comparison with data, a process known as {\it
    resummation}, discussed in section~\ref{sec:resummation}.
\item Real scattering events contain huge numbers of quarks and
  gluons, going way beyond what a perturbative QCD calculation can
  achieve. One must therefore estimate the effect of this additional
  radiation using well-motivated QFT-based arguments. We return to
  this in section~\ref{sec:experiment}.
\item Real scattering events do not contain free (anti-)quarks and
  gluons, due to the confinement property of QCD. They instead contain
  hadrons, and we must try and estimate how. This is also discussed in
  section~\ref{sec:experiment}.
\end{itemize}
This list -- which is not even complete -- is enough to scare off any
theorist who hopes that their research on scattering amplitudes might
have something to say about experiments. Let us thus retreat to the
relative safety of eq.~(\ref{cross-sec}), and get gradually closer to
what happens at the LHC.

\section{Cross-sections at hadron colliders}
\label{sec:partons}

The first thing we have to deal with is the issue of incoming
protons. At first glance, one might think that we could never
calculate cross-sections with incoming protons, given that -- with a
few exceptions in special circumstances or theories -- we only really
know how to calculate amplitudes using perturbation theory. However,
what saves us is the well-known property of {\it asymptotic
  freedom}~\cite{Gross:1973id,Politzer:1973fx}, according to which the
QCD expansion parameter $\alpha_s(\mu_R)$ becomes weaker with
increasing renormalisation scale $\mu_R$. Given that this is
identified with a typical energy scale $Q$ in the scattering process
of interest, we find that highly energetic protons will contain
approximately free (anti-)quarks and gluons. We can indeed calculate a
perturbative cross-section for the latter, and may then combine it
with appropriate distribution functions measuring how the
(anti-)quarks are distributed within the proton, in order to describe
the total proton-proton cross-section. This was first formulated by
Feynman~\cite{Feynman:1969wa,Feynman:1969ej}, who called it the {\it
  parton model}. At that time, partons were stipulated as hypothetical
elementary constituents of the proton, and were only later realised to
be the (anti-)quarks and gluons of QCD. The word ``parton'', however,
is still in use, as it is very convenient to have a single word that
can be used to denote any (anti-)quark or gluon!

More precisely, consider partons that emerge from the incoming
protons, where the latter can be taken to have momenta $P_1$ and
$P_2$. If the protons are very fast-moving, the emerging partons will
have momenta $\{p_i\}$ that are approximately collinear with the
$\{P_i\}$, and which may thus be written as
\begin{equation}
  p_i=x_i P_i,\quad 0\leq x_i\leq 1,
  \label{xidef}
\end{equation}
where no summation is implied on the right-hand side, and where $x_i$
represents the (longitudinal) momentum fraction of parton $i$. The
hadronic cross-section for a given process may then be written
as~\footnote{To simplify the notation in eq.~(\ref{parton}), we have
omitted the subscript on $\sigma$ that we used above to denote a
particular scattering process.}
\begin{equation}
  \sigma=\sum_{i,j\in\{q,\bar{q},g\}}\int_0^1 dx_1\int_0^1 dx_2\,
  f_i(x_1)f_j(x_2)\hat{\sigma}_{ij}(\{p_i\},\mu_R).
  \label{parton}
\end{equation}
Here $\hat{\sigma}_{ij}$ denotes the {\it partonic cross-section}, for
incoming (anti-)quarks and gluons. The quantity $f_i(x_j)$ is a {\it
  parton distribution function (PDF)}, that represents, loosely
speaking, the probability to find a parton with momentum fraction
$x_j$ inside incoming proton $i$. We must then sum over all possible
momentum fractions, and over all possible species of parton, including
over all possible flavours of quark $q$ and their anti-particles. The
latter might surprise you if you are familiar only with the
conventional undergraduate wisdom that protons contain two up quarks
and a down quark. Needless to say, this statement is not to be taken
too literally: in quantum field theory, partons are constantly popping
in and out of the vacuum due to virtual effects, so that the proton is
some sort of localised collection of quantum partonic froth, whose
{\it net} quantum numbers are the same as two up quarks and a down
quark! Whilst the partonic cross-section $\hat{\sigma}_{ij}$ is
calculable in perturbation theory, the parton distribution functions
$\{f_i\}$ are not. They can, however, be measured from experiment, and
used to predict the results of subsequent experiments.

\begin{figure}
  \begin{center}
    \scalebox{0.6}{\includegraphics{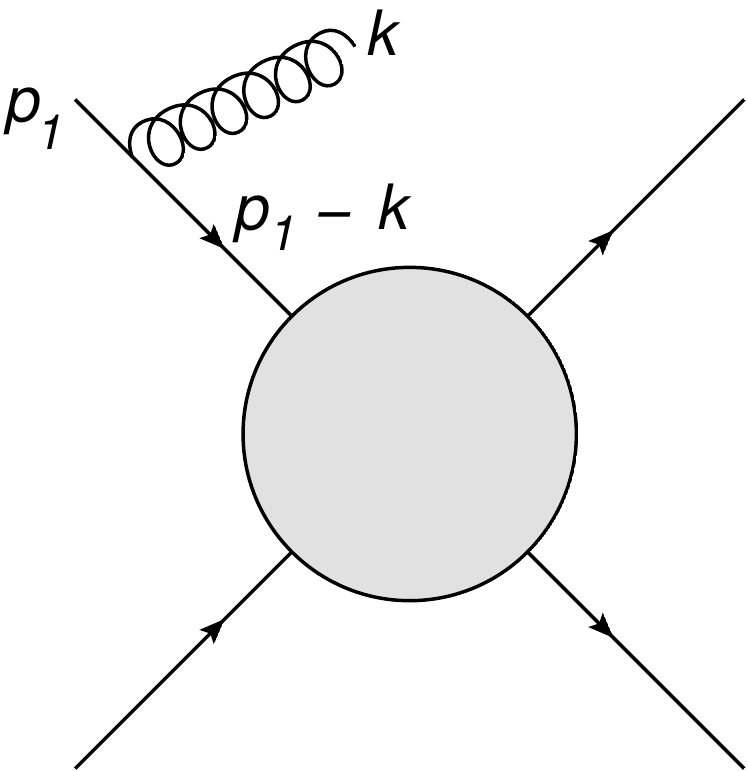}}
    \caption{Gluon radiation from an incoming quark leg.}
    \label{fig:radiation}
  \end{center}
\end{figure}

You may be wondering if the parton model can be rigorously justified
from first-principles QFT, where the relevant field theory is QCD in
this case. Indeed it can for certain scattering processes, using the
operator product expansion (see e.g. ref.~\cite{Collins:1984xc} for a
textbook treatment). More simply, we might ask what goes wrong if we
were to ignore eq.~(\ref{parton}), and try to use perturbation theory
to describe QCD scattering. At leading order (LO) in $\alpha_s$,
everything is well-behaved. However, at NLO and beyond, problems occur
due to emitting additional radiation from either incoming or outgoing
partons. An example is shown in figure~\ref{fig:radiation} which,
using Feynman rules (or otherwise) results in an extra factor in the
scattering amplitude for a given process
\begin{equation}
  \frac{1}{(p_1-k)^2}=\frac{1}{-2p_1\cdot k}
  =\frac{1}{-2|\vec{p}_1||\vec{k}|(1-\cos\theta)},
  \label{radfactor}
\end{equation}
where 4-momenta are labelled as in the figure, and we neglect the mass
of the incoming quark, given that this is negligible at all relevant
collider energies (at least for the up and down quarks). We have also
introduced the 3-momenta of the quark and gluon, $\vec{p}_1$ and
$\vec{k}$, as well as the angle $\theta$ between them. Clearly the
additional factor of eq.~(\ref{radfactor}) diverges in two cases: (i)
the additional radiation is {\it soft} ($|\vec{k}|\rightarrow 0$);
(ii) the additional radiation is {\it collinear} with the emitting
particle ($\theta\rightarrow 0$). Furthermore, these cases are not
mutually exclusive: the radiation may be soft {\it and} collinear,
which is even more divergent! These are known as {\it infrared (IR)}
divergences, to distinguish them from the UV divergences encountered
above. Although we considered the example of real radiation here, the
virtual (loop) corrections to the amplitude are also IR-divergent, and
the solution to this problem is well-known. First, one must be careful
to consider only those cross-sections and related observables that are
{\it infrared-safe}, meaning that the definition is robust under the
inclusion of additional soft and / or collinear radiation. For
example, the cross-section for a single Higgs boson accompanied by any
amount of QCD radiation is IR-safe, whereas the cross-section for a
Higgs boson plus exactly three detected gluons is not. Once we have
chosen an IR-safe observable, the divergences are guaranteed to cancel
in QED by the {\it Bloch-Nordsieck theorem}~\cite{Bloch:1937pw}, if we
add together all virtual and real contributions at any given order in
perturbation theory. In QCD, the {\it KLN
  theorem}~\cite{Kinoshita:1962ur,Lee:1964is} theorem states that this
cancellation will only occur if we include initial states containing
arbitrary numbers of incoming particles, which is inconvenient for
describing scattering processes with two incoming beams. Instead
requiring only two incoming particles leads to cancellation of all
soft singularities, together with collinear singularities associated
with the outgoing particles in the final state. However, one is left
with uncancelled collinear singularities associated with the two
incoming particles, which potentially pose a serious problem.

What saves us is the presence of the parton distributions $\{f_i\}$ in
eq.~(\ref{parton}). We are clearly free to redefine both the partons
and the partonic cross-section such that eq.~(\ref{parton}) remains
invariant. Thus, we can choose to remove collinear divergences from
the partonic cross-section by absorbing them into the parton
distributions. In other words, one replaces the {\it bare parton
  distributions} appearing in eq.~(\ref{parton}) with modified ones,
chosen so that singularities in the partonic cross-section are
removed. One way to do this is simply to dictate that all radiation
below a particular {\it factorisation scale} $\mu_F$ is reabsorbed
into a redefined parton distribution, so that it is absent from the
partonic cross-section. However, this procedure is typically not used
in practical calculations, given that applying momentum cut-offs is
not gauge invariant. Instead, dimensional regularisation is often used
in $d=4-2\epsilon$ dimensions, in which IR divergences show up as
poles in $\epsilon$. The factorisation scale then emerges through the
additional scale $\mu$ that is introduced to keep the coupling
dimensionless. A particular means of removing divergences from the
partonic cross-section is called a {\it factorisation scheme} and,
whilst the partonic cross-section becomes finite, the calculated
parton distributions are not. This is not a problem: we do not claim
to be able to calculate the partons in perturbation theory, and so
instead can simply measure them from experiment. Indeed, this is
reminiscent with what happens when we remove UV singularities via
renormalisation: in doing so, we lose the ability to calculate the
coupling constant, which instead becomes a measurable parameter in
some given scheme.

The upshot of the above discussion is that eq.~(\ref{parton2}) gets replaced
by the more general formula
\begin{eqnarray}
  \sigma&=\sum_{i,j\in\{q,\bar{q},g\}}\int_0^1 dx_1\int_0^1 dx_2\,
  f_i(x_1,\mu_F)f_j(x_2,\mu_F)\hat{\sigma}_{ij}(\{p_i\},\mu_R,\mu_F)
  \nonumber\\
  &\quad +{\cal O}\left(\frac{\Lambda^2}{Q^2}\right).
  \label{parton2}
\end{eqnarray}
That is, both the partons and partonic cross-section now depend on the
factorisation scale $\mu_F$, which typically enters through
dimensionless ratios involving an energy scale $Q$ associated with the
scattering process. As for the renormalisation scale $\mu_R$,
dependence on $\mu_F$ cancels at a given order in perturbation theory,
but there can be a residual dependence associated with missing
higher-order corrections, and this takes the form of logarithms of
ratios of the factorisation scale with typical energy scales involved
in the scattering process.  As for the case of $\mu_R$ above, one can
reduce this dependence by choosing $\mu_F$ to be such a typical energy
scale $Q$, and there is clearly some ambiguity in this choice. Varying
$\mu_R$ and $\mu_F$ around some default value then gives a measure of
the {\it theoretical uncertainty} of a given cross-section
prediction. We have also noted in eq.~(\ref{parton2}) that the parton
model formula receives formal corrections when derived from
first-principles QFT, involving ratios of the QCD confinement scale
$\Lambda$ and the typical energy scale $Q$. Provided that $Q$ is large
enough that we trust perturbation theory in the first place, these
so-called {\it power corrections} will be small.

\section{Fixed-order perturbation theory}
\label{sec:fixedorder}

So far the job of a theorist seems clear enough: to compare theory to
data at a hadron collider such as the LHC, we must calculate
amplitudes involving incoming partons (i.e. (anti-)quarks and gluons),
turn these into partonic cross-sections using eq.~(\ref{cross-sec}),
and then combine these with parton distributions using one of the many
versions of these which are available (see
e.g. refs.~\cite{Bailey:2020ooq,NNPDF:2019ubu,Alekhin:2018pai,Hou:2019efy}
for recent analyses). However, anyone who has tried to calculate
scattering amplitudes for relevant scattering processes knows that
this becomes extraordinarily difficult as the number of loops or
number of legs increases. Traditionally, QCD amplitudes were
calculated using Feynman rules and diagrams, which represent the
$L$-loop amplitude as a sum of integrals of the form
\begin{equation}
  {\cal A}^{(L)}(\{p_i\})=\sum_{{\rm diags}\,\, j} \left(\prod_{k=1}^L
  \int \frac{d^d l_k}{(2\pi)^d}\right)\frac{c_j\,
    {\cal N}_j(\{p_i\},\{l_k\})}{{\cal D}_j(\{p_i\},\{l_k\})}.
  \label{feynmanint}
\end{equation}
Here each individual diagram $j$ has a colour factor $c_j$, and
integrals over the $L$ loop momenta $\{l_k\}$. There is then a
kinematic numerator ${\cal N}_j$ and denominator ${\cal D}_j$, each of
which depends on both the external momenta $\{p_i\}$ and loop momenta
$\{l_k\}$ in general. Terms in the integrand with zero, one or more
than one powers of loop momenta in the numerator are referred to as
{\it scalar}, {\it vector} or {\it tensor} integrals respectively, and
it is in fact possible to apply algebraic identities to reduce all
higher-rank integrals to scalar ones. This may be formalised in a
process known as {\it Passarino-Veltman
  reduction}~\cite{Passarino:1978jh}, which may be automated in
principle, albeit with some subtle caveats for higher-leg
processes. Then, all $L$-loop amplitudes can in principle be
straightforwardly calculated once a suitable basis for all $L$-loop
scalar integrals is known, a problem which is entirely solved for
$L=1$ (see e.g. ref.~\cite{Ellis:2007qk}). Roughly speaking, most QCD
calculations up until the mid-2000s were calculated using this
approach, which ultimately proved to be unsustainable, not least due
to the fact that the number of Feynman diagrams grows factorially as
the order in perturbation theory increases. 

More modern methods for calculating amplitudes rely on the observation
that, once a basis of $L$-loop scalar integrals is known, one can
simply expand a given amplitude in terms of this basis, and then
develop clever methods for fixing the coefficients that bypass Feynman
diagrams altogether. A heavily-used method is that of {\it generalised
  unitarity}~\cite{Bern:1994zx,Bern:1994cg,Bern:1995db,Bern:1997sc,Britto:2004nc,Brandhuber:2005jw},
which involves first classifying the list of scalar integrals
according to how many internal lines (or {\it propagators}) they
have. Next, one may perform {\it unitarity cuts}, consisting of
placing one or more propagators on-shell. If one takes the maximal
number of such cuts at a given loop order, only the integrals with the
maximal number of internal lines survive, so that one may
straightforwardly fix their coefficients. One may then iterate the
procedure to fix the integrals with one less than the maximal number
of internal lines, and so on. The so-called {\it cut-constructible}
part of the amplitude thus obtained must then be supplemented by an
additional {\it rational piece}, for which various methods exist
~\cite{Britto:2004ap,Bern:2005hs,Bern:2005ji,Badger:2005zh,Badger:2005jv,Anastasiou:2006jv,Anastasiou:2006gt,Giele:2008ve,Huang:2013vha,Giele:2008ve}
(see e.g. ref.~\cite{Ellis:2011cr} for a comprehensive
review). Ultimately, the effect of such tools is to reduce the
computational complexity of multiloop calculations from being
factorial in the perturbative order, to being merely
polynomial. Indeed, the calculation of one-loop processes in QCD has
now been automated, such that general purpose computer programs exist
that allow users to specify a given initial and final state (see
e.g. refs.~\cite{Gleisberg:2007md,Hirschi:2011pa,Bellm:2015jjp}, or
refs.~\cite{Sherpa:2019gpd,Alwall:2014hca,Bellm:2019zci} for their
latest incarnations).  One then presses a button, waits for a
reasonable (but not too inconvenient) length of time, and receives
results for (differential) cross-sections!

At two-loop level and beyond, a full basis of scalar integrals is not
known, and discussion is still ongoing regarding the best way to
systematise calculations. One can of course focus on specific
scattering processes rather than trying to make general statements.
Then a reduced set of integrals appear, but the difficult problem
remains of carrying out the integrals themselves, in terms of known
functions. In recent years, this has generated a fascinating dialogue
between pure mathematicians and theoretical physicists, and more
details can be found in
refs.~\cite{Abreu:2022mfk,Blumlein:2022zkr,Papathanasiou:2022lan}
(chapters 3, 4 and 5 of this review~\cite{Travaglini:2022uwo}). What
is clear, however, is that there is a clear direction of travel of new
computational techniques on average. They tend to be developed first
in a formal \texttt{hep-th} context, typically due to the fact that it
is easier to probe new structures in highly symmetric theories such as
${\cal N}=4$ Super-Yang-Mills. However, these same techniques then
filter down to more immediately applicable theories such as QCD, and
become a standard part of the {\texttt hep-ph} toolkit. Very often,
individual researchers will be working in both subfields, providing a
vibrant counterpoint to the once-prevalent notion that ``theory'' and
``phenomenology'' are distinct activities, whose practitioners are not
able -- or unwilling -- to communicate with each other.

The calculation of a given scattering process at higher orders may
proceed completely analytically, or (partially) numerically
(e.g. numerical integration of scalar integrals may be used). This may
seem distasteful to those formal theorists who are used to being able
to fully probe the analytic structure of their results. But numerical
results are ultimately what is needed anyway to compare with
experiment, and numerical methods have allowed us to carry out
calculations that otherwise would simply not have been
possible. However, the use of numerical methods raises a significant
issue: as discussed above, the calculation of physical observables
involves infrared singularities at intermediate stages. Although these
cancel in final results, and may be regulated in separate parts of a
calculation (e.g. cross-section contributions involving different
numbers of loops), one cannot combine numerical results for
individually IR-divergent parts of an observable, due to the large
numerical uncertainties that result. One must thus devise a suitable
{\it infrared-subtraction scheme} for organising the perturbative
calculation, such that large numerical cancellations never
appear. This problem was solved a long time ago at one-loop (for
popular schemes, see refs.~\cite{Frixione:1995ms,Catani:1996vz}), and
the frontier is how to systematically accomplish this at two-loop
level and
beyond~\cite{Gehrmann-DeRidder:2005btv,Currie:2013vh,Czakon:2010td,Czakon:2011ve,Boughezal:2011jf,Anastasiou:2003gr,Binoth:2004jv,Catani:2007vq,Boughezal:2015eha,Gaunt:2015pea,DelDuca:2015zqa,Caola:2017dug,Delto:2019asp,Caola:2019nzf,Magnea:2018hab,Magnea:2018ebr,Caola:2019pfz,Asteriadis:2019dte,Cacciari:2015jma}.

The motivation of a phenomenologist can be very different to a more
formal theorist: the choice of which amplitude or scattering process
to consider next is much more likely to be guided by experiment. More
specifically, members of the two large general-purpose search
experiments at the LHC (ATLAS and CMS) have often produced wishlists
of scattering processes for which they would like higher-order
corrections. A recent example can be found in
ref.~\cite{Huss:2022ful}, which also reviews state-of-the-art
calculations tools. Various processes are discussed, and requests made
for both QCD and electro-weak (EW) corrections. The interplay between
including both QCD and EW perturbative information needs some careful
accountancy (see ref.~\cite{Denner:2019vbn} for a
review). Furthermore, differential cross-sections are needed as well
as total cross-sections. These typically lag behind total
cross-sections in terms of the available precision: the presence of
more resolved momenta in the final state means that there are more
energy scales in the problem, which complicates the computation of
relevant loop integrals, as well as the general book-keeping of the
calculation itself.

The absolute cutting-edge in fixed-order perturbation theory is
N$^3$LO~\cite{Anastasiou:2015vya,Anastasiou:2016cez,Dreyer:2016oyx,Dreyer:2018qbw,Duhr:2020seh},
and the vast majority of interesting processes are still known only at
NLO. For total cross-sections, only a few orders in perturbation
theory are typically needed in order to start to approach a
theoretical uncertainty of sub-percent level (as estimated by
variation of the factorisation and renormalisation scales, whose
residual dependence tells us about missing higher-order
corrections). However, for differential quantities serious problems
can occur, in that the perturbation expansion becomes unstable. We
explore this in the following section.

\section{Resummation}
\label{sec:resummation}

\begin{figure}
  \begin{center}
    \scalebox{0.8}{\includegraphics{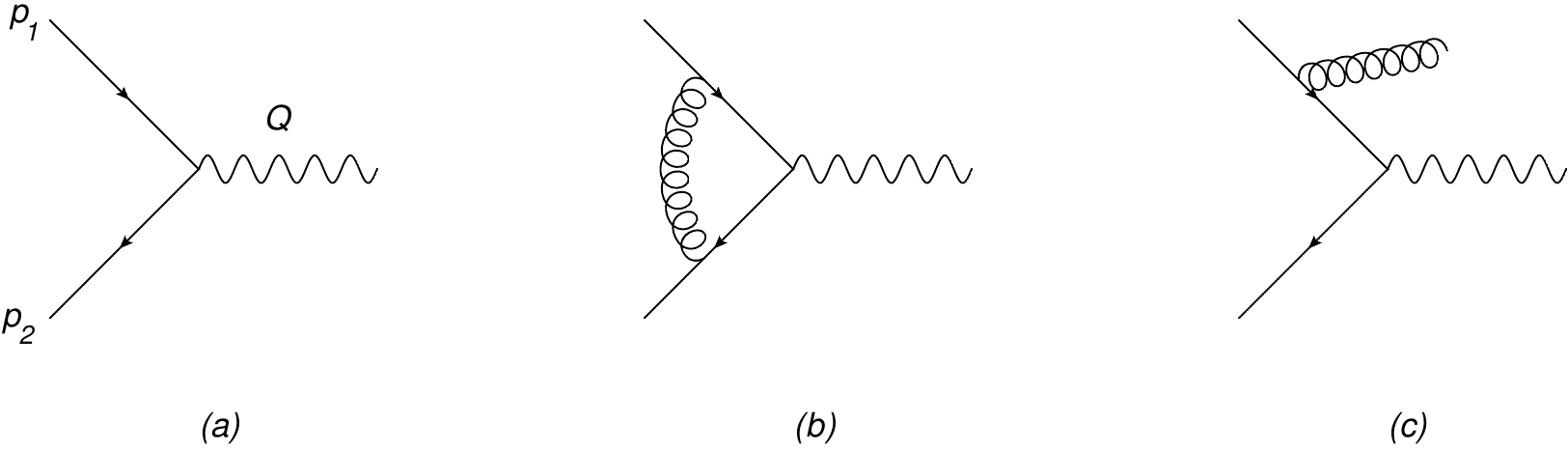}}
    \caption{(a) LO Feynman diagram for the production of an off-shell
      vector boson by a quark-antiquark pair; (b) a virtual correction
      at NLO; (c) a real correction at NLO.}
    \label{fig:DY}
  \end{center}
\end{figure}

For a classic example of how perturbation theory can go awry, let us
consider the {\it Drell-Yan process}, in which an off-shell photon is
produced that eventually decays to a lepton pair. The LO Feynman
diagram is shown in figure~\ref{fig:DY}(a), where we label 4-momenta
as shown. It is then conventional to define the variable
\begin{equation}
  z=\frac{Q^2}{s},
  \label{zdef}
\end{equation}
which may be loosely interpreted as the fraction of the partonic
centre of mass energy $s$ that is carried by the photon (of virtuality
$Q^2$).
The LO differential partonic cross-section in this variable then has
the form
\begin{equation}
  \frac{d{\hat{\sigma}}}{dz}=\sigma_0\,\delta(1-z),
  \label{DYLO}
\end{equation}
for some $\sigma_0$, where the delta function reflects the fact that
the photon is carrying all of the energy in the final state at LO, so
that we must have $z=1$. In computing high-order corrections, we must
define the total cross-section for Drell-Yan production to include any
amount of additional QCD radiation, in line with our earlier comments
regarding infrared safety. We must then include both virtual and real
corrections, examples of which are shown in figure~\ref{fig:DY}(b) and
(c). These are individually infrared divergent, such that any formal
singularity cancels upon combining all real and virtual graphs (bar
those initial-state collinear singularities that must be absorbed into
the parton distributions). However, upon combining all contributions,
the NLO differential cross-section for the $q\bar{q}$ initial state
turns out to have the following form:
\begin{eqnarray}
  \frac{d\hat{\sigma}_{q\bar{q}}^{(1)}}{dz}&= \frac{\alpha_s
    C_F}{2\pi}\left[4(1+z^2)\left(\frac{\log(1-z)}{1-z}\right)_+
    -2\frac{1+z^2}{1-z}\log(z)\right.\nonumber\\ &\left.\quad+\delta(1-z)\left(\frac{2\pi^2}{3}-8\right)
    \right].
  \label{qqNLO}
\end{eqnarray}
where $C_F$ is a colour-dependent constant, and the $+$ notation
denotes a so-called {\it plus distribution}, defined by its action on
a test function $g(z)$ as:
\begin{equation}
  \int_0^1 dz f_+(z) g(z)\equiv \int_0^1 dz f(z)[g(z)-g(1)].
  \label{plusdef}
\end{equation}
We now have a non-trivial dependence on the variable $0\leq z\leq 1$,
reflecting the fact that additional radiation may carry away some of
the energy in the final state. However, we also see that
eq.~(\ref{qqNLO}) contains a term that is highly divergent as
$z\rightarrow 1$, involving a logarithm of $\xi\equiv (1-z)$ divided
by $\xi$ itself. Finiteness of the total cross-section is guaranteed
by eq.~(\ref{plusdef}), but it is still the case that the logarithmic
term becomes extremely large near $z\rightarrow 1$. This threatens the
validity of perturbation theory: expansion in the coupling $\alpha_s$
only makes sense if the coefficients of the expansion are small. At
the very least, we should look at higher orders in perturbation theory
and see how the coefficients behave as $z\rightarrow
1$. Unfortunately, however, the problem gets even worse!  At NNLO, for
example, we have
\begin{equation}
  \frac{d\sigma_{q\bar{q}}^{(2)}}{dz}=\left(\frac{\alpha_s C_F}{2\pi}\right)^2
  \left[128\left(\frac{\log^3(1-z)}{1-z}\right)_+-256\left(
    \frac{\log(1-z)}{1-z}\right)_++\ldots\right],
  \label{qqNNLO}
\end{equation}
where the ellipsis denotes terms that are suppressed by a power of
$(1-z)$. The highest power of the log in $(1-z)$ has gotten larger,
and one may indeed show that this pattern recurs at higher orders, so
that perturbation theory is indeed breaking down.

To understand the origin of the problem, note that $z\rightarrow 1$
amounts to the photon carrying all the final state energy, so that any
additional radiation is forced to be soft. Although the formal soft
divergences cancelled when we added together real and virtual
contributions, the large logarithms are ``echoes'' of the fact that
these singularities were present in the first place. Here we have
considered the case where additional radiation is forced to be
soft. In other scattering processes, it may also be the case that the
radiation is forced to be collinear in some kinematic regions, and
this also ends up leading to large logs. Indeed, the generic nature of
this explanation suggests the following general remarks. For processes
involving heavy particles being produced near threshold (i.e. with
only just enough 4-momentum), we can define a {\it threshold variable}
$\xi$, such that $\xi\rightarrow0$ near threshold (in our above
example, we had $\xi=(1-z$)). Then, the general structure of the
partonic differential cross-section in $\xi$ can be shown to be
\begin{equation}
  \frac{d\hat{\sigma}}{d\xi}=\sum_{n=0}^\infty \alpha_s^n
  \sum_{m=0}^{2n-1}\left[
    c_{nm}^{(0)}\left(\frac{\log^m\xi}{\xi}\right)_+
    +c_{nm}^{(1)}\log^m\xi+\ldots\right].
  \label{sigmaxi}
\end{equation}
The first set of terms on the right-hand side generalise the large
terms we have already seen in eqs.~(\ref{qqNLO}, \ref{qqNNLO}). They
are called {\it leading power (LP)} threshold logs, and correspond to
the emission of purely soft and / or collinear radiation. The
remaining terms constitute a systematic expansion in $\xi$, such that
the second set of terms comprises the {\it next-to-leading power
  (NLP)} contributions. The ellipsis then denotes terms which are NNLP
and beyond. It is then genuinely true that we cannot trust
perturbation theory as $\xi\rightarrow 0$. Furthermore, this is not
just a problem in principle, but very much one in practice: there are
many observables at the LHC where the instability of perturbation
theory becomes a problem. 

The solution sounds impossible at first. We must somehow work out what
the large logarithms are to all orders in $\alpha_s$, and sum them up
to get a function of $\alpha_s$ that is much better behaved than any
fixed-order perturbation expansion. You will be familiar with this
idea from your undergraduate days. Consider, for example, the toy
function
\begin{equation}
  e^{-\alpha_s x}=\sum_{n=0}^\infty \frac{\alpha_s^n(-x)^n}{n!}.
  \label{expalpha}
\end{equation}
Each term on the right-hand side diverges as $x\rightarrow\infty$, but
the left-hand side is perfectly well-behaved. What's more, this
example is not quite as trivial as it might first appear: when the
threshold logarithms in eq.~(\ref{sigmaxi}) are summed up, they do
indeed tend to exponentiate. What makes this possible is the fact that
we can understand the large logs as being related to infrared
divergences. The latter, as is well-known, can be classified to all
orders in perturbation theory (see e.g. ref.~\cite{Agarwal:2021ais}
for a review of this subject), so that summing threshold logs in
perturbation theory amounts to classifying the structure of infrared
singularities.

The process of summing up the large logs is known as {\it
  resummation}, and typically proceeds as follows. First, one can take
the most divergent logs at each order in $\alpha_s$, which at leading
power go as $\alpha_s^n \log^{2n-1}\xi/\xi$. These are referred to as
{\it leading logarithmic (LL)} terms, and are the easiest to
resum. Next, one may consider the {\it next-to-leading logarithmic
  (NLL)} terms $\sim \alpha_s^n \log^{2n-2}\xi/\xi$ to all orders, and
so on. This expansion implies a reordering of perturbation theory, and
it is this that the prefix ``re-'' in ``resummation'' is meant to
signify. There are by now many different approaches to resumming logs
in QCD, involving diagrammatic
arguments~\cite{Parisi:1979xd,Curci:1979am,Sterman:1986aj,Catani:1989ne,Gatheral:1983cz,Frenkel:1984pz,Sterman:1981jc},
use of Wilson lines~\cite{Korchemsky:1992xv,Korchemsky:1993uz},
renormalisation group arguments~\cite{Forte:2002ni}, and effective
field
theory~\cite{Becher:2006nr,Schwartz:2007ib,Bauer:2008dt,Chiu:2009mg}
(see refs.~\cite{Luisoni:2015xha,Becher:2014oda,Campbell:2017hsr} for
pedgagogical reviews). Common to all of them, however, is the fact
that soft and collinear radiation essentially factorises from the
underlying hard scattering process. That is, a general $n$-point
amplitude dressed by soft and collinear radiation has the following
form (see e.g. ref.~\cite{Dixon:2008gr}):
\begin{equation}
  {\cal A}_n={\cal H}_n\,{\cal S}\,\frac{\prod_{i=1}^n J_i}
  {\prod_{i=1}^n {\cal J}_i}.
  \label{softcolfac}
\end{equation}
Here ${\cal H}_n$ is a so-called {\it hard function}. It depends on
the particular scattering process, but is infrared finite. The {\it
  soft function} ${\cal S}$ collects all singularities associated with
purely soft radiation, and the {\it jet functions} $J_i$ collect
collinear singularities associated with external leg $i$. However,
radiation can be both soft and collinear, which means it has been
counted twice by being included in both the soft and jet
functions. Thus, one must divide by {\it eikonal jet functions} ${\cal
  J}_i$ to remove the double-counting. It turns out that the soft and
jet functions have universal definitions, which are independent of the
particular scattering process being considered. The physics of this is
indeed straightforward: soft radiation has vanishing momentum, and
collinear radiation has vanishing momentum transverse to the relevant
particle direction. In both cases, the emitted radiation thus has an
infinite Compton wavelength, and cannot resolve the details of the
underlying hard scattering process. We then expect it to factorise,
leading to something like eq.~(\ref{softcolfac}).

Studying the soft function in more detail allows us to make contact
with more formal amplitudes literature. In QCD, one may define it as a
vacuum expectation value of Wilson line operators. That is, if
$\{\beta_i\}$ are the 4-velocities of the coloured external particles,
one may write
\begin{equation}
  {\cal S}(\{\beta_i\})=\langle 0|\Phi_1\,\Phi_2\,\ldots \Phi_n|0\rangle,
  \label{Sdef}
\end{equation}
where
\begin{equation}
  \Phi_i={\cal P}\exp\left[i g_s \int_{x_i=\tau_i\beta_i} dx_i^\mu
    A_\mu^a {\bf T}^a_i \right]
  \label{Phidef}
\end{equation}
is the usual gauge theory Wilson line consisting of the integral of
the gauge field along a curve, which has been taken in each case to be
the classical straight-line trajectory $x_i=\tau_i\beta_i$ ($0\leq
\tau_i\leq \infty$) associated with the 4-velocity $\beta_i$. Also
${\bf T}^a_i$ is a colour generator associated with external line $i$,
and ${\cal P}$ denotes path-ordering of the colour generators along
the contour. One way of seeing why eq.~(\ref{Sdef}) is correct is to
note that each of the Wilson line factors contains a coupling of the
gauge field to each coloured external particle. By Fourier
transforming the exponent of eq.~(\ref{Phidef}) to momentum space, we
can then interpret the Feynman rule for this coupling. To this end, we
may rewrite the exponent of eq.~(\ref{Phidef}) as
\begin{displaymath}
  ig_s {\bf T}_i^a \int dx_i^\mu \int\frac{d^d k}{(2\pi)^d}
  \tilde{A}_\mu^a(k)e^{ik\cdot x_i}=
  \int\frac{d^d k}{(2\pi)^d}\tilde{A}_\mu^a(k)\left[ig_s{\bf T}^a_i
    \beta_i^\mu \int_0^\infty d\tau_i e^{i(k\cdot \beta_i)\tau_i} 
    \right].
\end{displaymath}
It is straightforward to carry out the integral over the parameter
$\tau_i$, and we obtain\footnote{The upper limit of the integral in
$\tau_i$ vanishes once the Feynman $i\varepsilon$ prescription is
properly taken into account.}
\begin{equation}
  \int\frac{d^d k}{(2\pi)^d}\tilde{A}_\mu^a(k)\left[-g_s{\bf T}^a_i
    \frac{\beta_i^\mu}{\beta_i\cdot k}\right].
  \label{feynmanrule}
\end{equation}
The square bracket contains the {\it eikonal Feynman rule} for the
emission of soft radiation, that enters the well-known {\it soft
  theorem} in gauge theory~\cite{Weinberg:1965nx}, describing the
emission of soft radiation. The above argument is somewhat technical,
but there is a more physical way to see that the soft function should
be given by eq.~(\ref{Sdef}). If the coloured external particles are
emitting purely soft radiation, they cannot recoil against anything,
and so by definition must be following classical straight-line
trajectories. Then, the only quantum behaviour they are allowed to
have is to experience a phase change which, if it is to have the right
gauge-covariance properties to form part of a scattering amplitude,
must be described by a Wilson line operator, which is known to
transform covariantly.

We see, then, that classifying the structure of infrared singularities
in scattering amplitudes amounts to studying vacuum expectation values
of Wilson lines. Computations involving Wilson lines occur widely
throughout the formal amplitudes literature, not least due to the
well-known duality between scattering amplitudes in ${\cal N}=4$
Super-Yang-Mills theory, and certain Wilson loops formed from the
particle momenta~\cite{Alday:2007hr}. A key property of Wilson lines
that we need for our present purposes is that eq.~(\ref{Sdef}) is
subject to UV singularities, associated with the cusp at which the
Wilson lines meet. These then correspond to the IR singularities of
the original amplitude. To see how this works, consider some external
particles that are emitting virtual radiation, as in
e.g. figure~\ref{fig:DY}(b). Relative to leading order, there will be an additional propagator on the upper line
\begin{displaymath}
  \sim\frac{1}{(p_1-k)^2}=\frac{1}{-2p_1\cdot k+ k^2},
\end{displaymath}
where $k$ is the 4-momentum of the exchanged gluon. In the soft limit,
one linearises the propagator by neglecting the term of ${\cal
  O}(k^2)$, leading to the eikonal Feynman rule dependence of
eq.~(\ref{feynmanrule}). However, $k$ is a loop momentum, and thus
replacing the denominator in this fashion changes the behaviour of the
integrand as $k$ becomes large. Taking also the lower line into
account, one modifies the denominators in the loop integral over $k$
as follows:
\begin{equation}
  \int\frac{d^d k}{(2\pi)^d}\frac{1}{k^2(-2p_1\cdot k+k^2)(2p_2\cdot k+k^2)}
  \rightarrow
  \int\frac{d^d k}{(2\pi)^d}\frac{1}{k^2(-2p_1\cdot k)(2p_2\cdot k)}.
  \label{intmod}
\end{equation}
The original integral is UV-finite, but the modified integral is
logarithmically divergent. Hence, taking the soft approximation has
introduced a {\it spurious} UV divergence. Furthermore, the second
integral turns out to be scaleless in dimensional regularisation, and
thus formally vanishes. Thus, the spurious UV singularity precisely
cancels the original IR singularity we are interested in. It follows
that the UV singularity of a Wilson line integral (involving the
linear propagators) matches the IR singularity of the amplitude, and
this property in fact generalises to all orders in the soft function.

\begin{figure}
\begin{center}
\scalebox{0.5}{\includegraphics{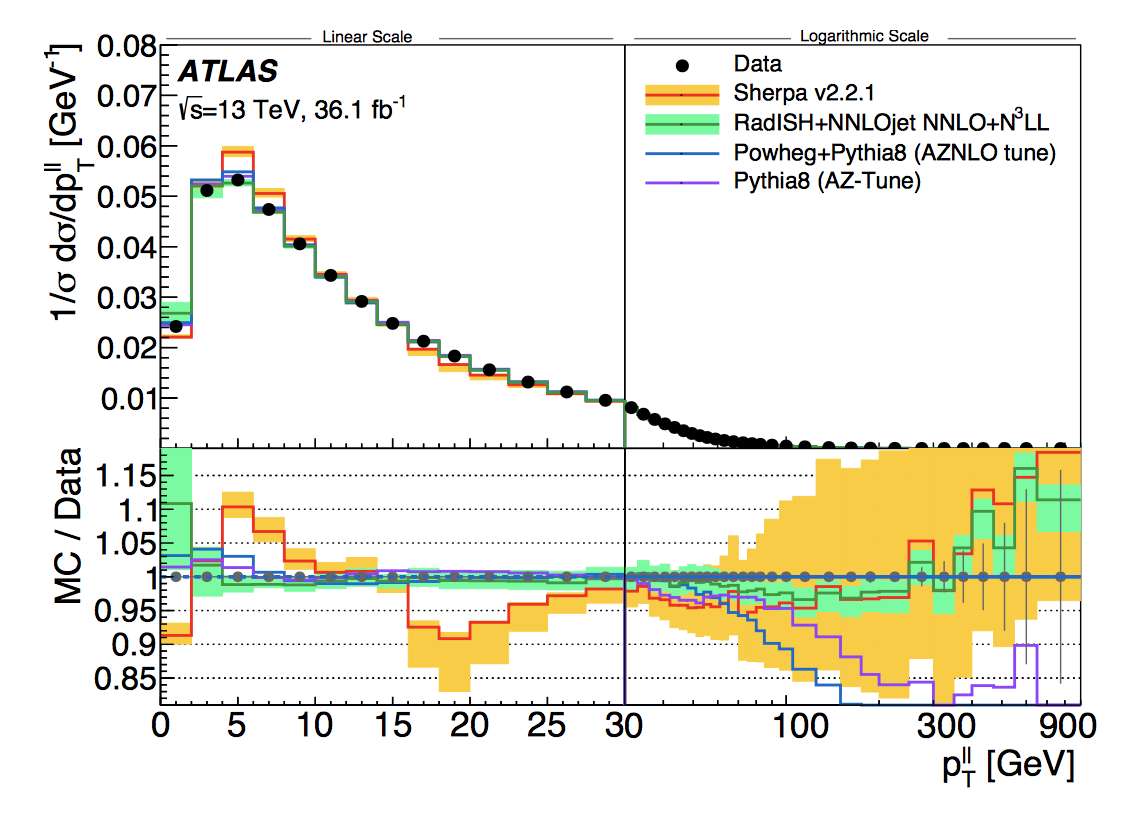}}
\caption{The distribution of the transverse momentum of a lepton pair
  in Drell-Yan production. The curve labelled with ``NNLO+N$^3$LL''
  includes threshold lograithms resummed to all orders in perturbation
  theory.}
\label{fig:DYPT}
\end{center}
\end{figure}

The UV singularities of VEVs of Wilson lines are controlled by a
quantity called the {\it soft anomalous dimension}, otherwise known as
the {\it cusp anomalous dimension} when only two Wilson lines are
involved (again, see ref.~\cite{Agarwal:2021ais} for a review). Its
specific form depends upon the particular theory being considered, and
whether there are only two Wilson lines, or many. From the above
discussion, it follows that knowledge of the soft anomalous dimension
is a key ingredient in being able to perform resummation: (i) the soft
anomalous dimension determines the UV singularities of Wilson lines;
(ii) these in turn are directly related to the IR singularities of
amplitudes; (iii) classifying the latter allows us to resum large
logarithms to all-orders in perturbation theory, thus getting sensible
results for certain observables, that we can compare with
experiments. Further anomalous dimensions control the behaviour of
collinear singularities, but we stress the soft anomalous dimension
here given that it is a widely studied quantity in a variety of formal
contexts, where the people involved may not have realised its role in
helping us to get the most out of LHC data: each successive order in
the soft anomalous dimension allows us to sum up a further tower of
large logs (NLL, NNLL, NNNLL...) in cross-sections of interest. The
state of the art in QCD is three-loop order for processes involving
only two coloured particles at
LO~\cite{Polyakov:1980ca,Arefeva:1980zd,Dotsenko:1979wb,Brandt:1981kf,Korchemsky:1985xj,Korchemsky:1985xu,Korchemsky:1987wg,Grozin:2015kna,Grozin:2014hna}
(four loops in QED~\cite{Bruser:2020bsh}). For many Wilson lines, the
soft anomalous dimension is known at three-loop order
\cite{Korchemsky:1993hr,Korchemskaya:1996je,Korchemskaya:1994qp,Catani:1996vz,Catani:1998bh,Sterman:2002qn,Dixon:2008gr,Kidonakis:1998nf,Bonciani:2003nt,Dokshitzer:2005ig,Aybat:2006mz,Gardi:2009qi,Becher:2009cu,Becher:2009qa,Gardi:2009zv,
  Dixon:2009gx,Dixon:2009ur,DelDuca:2011ae,Caron-Huot:2013fea,Ahrens:2012qz,Naculich:2013xa,Erdogan:2014gha,Gehrmann:2010ue,Agarwal:2021zft,Almelid:2015jia,Gardi:2016ttq}
if the Wilson lines are lightlike, as would be relevant for the
scattering of massless particles. For non-lightlike Wilson lines
(which would be relevant for e.g. top quark pair production), the
state-of-the-art is two-loop
order~\cite{Kidonakis:2009ev,Mitov:2009sv,Becher:2009kw,Beneke:2009rj,Czakon:2009zw,Ferroglia:2009ii,Chiu:2009mg,Mitov:2010xw,Gardi:2013saa,Falcioni:2014pka,Henn:2013tua},
although progress towards a full three-loop result has been
reported~\cite{Gardi:2010rn,Gardi:2011wa,Dukes:2013wa,Gardi:2013ita,Falcioni:2014pka,Gardi:2021gzz,Mitov:2010rp,Vladimirov:2015fea}. Interestingly,
the three-loop soft anomalous dimension for massless particles has a
remarkably simple analytic form. It was subsequently shown that it
could be obtained without explicit calculation, by using a {\it
  bootstrap approach}, in which one expands it in a basis of known
functions, before applying known constraints from collinear and high
energy limits to fix the coefficients~\cite{Almelid:2017qju}. This
strongly suggests that further techniques from formal amplitudes
research -- involving both the theory of special functions, and
insights into which functions can appear at which loop orders -- may
prove to be highly useful in extending our ability to resum
perturbation theory. To illustrate the importance of this goal,
figure~\ref{fig:DYPT} shows the measured distribution of the
transverse momentum (relative to the incoming beams) of the virtual
particle produced in Drell-Yan production, and which decays to a pair
of leptons. At LO in perturbation theory, the Feynman diagram of
figure~\ref{fig:DY}(a) tells us that the transverse momentum must be
exactly zero, as there is nothing for the virtual photon to recoil
against. At NLO, it may recoil against an emitted gluon, but the
collinear singularity of this emission means that the distribution
would diverge as the transverse momentum goes to zero, due to a large
logarithm involving the transverse momentum. Only by resumming large
logs to all orders does the theory prediction match the data. This is
one of probably hundreds of individual observables at the LHC where
resummation is important. Also, desired improvements in the order of
the logarithms summed in specific processes form part of experimental
wishlists~\cite{Huss:2022ful}.\\
As well as the order of the logs, another potential improvement is to
include the terms that we have so far neglected in
eq.~(\ref{sigmaxi}). In particular, whilst a lot is known about the
leading power (LP) terms, much less is known about the next-to-leading
power (NLP) terms. If the LP terms are dictated by the emission of
soft and collinear radiation, the NLP terms are in principle described
by {\it next-to-soft} (or next-to-collinear)
radiation. Coincidentally, this is also a topic that has been widely
studied on \texttt{hep-th} in recent years, due in particular to the
fact that next-to-soft properties of amplitudes have been linked to
symmetries at asymptotic
infinity~\cite{Cachazo:2014fwa,Casali:2014xpa,Strominger:2017zoo},
thus starting an ongoing programme known as {\it celestial
  holography}, discussed in chapter 11 of this
review~\cite{McLoughlin:2022ljp}. As for the use of Wilson lines
discussed above, practitioners in this field may be entirely unaware
that the need to understand next-to-soft physics has a highly
practical application at the LHC. The need to describe and potentially
resum the NLP terms in eq.~(\ref{sigmaxi}) has been emphasised in
e.g. refs.~\cite{Kramer:1996iq,Ball:2013bra,Bonvini:2014qga,vanBeekveld:2019cks,vanBeekveld:2021hhv}. Indeed,
the study of such corrections has a remarkably long history, beginning
with the classic work of refs.~\cite{Low:1958sn,Burnett:1967km} for
massive external particles, which was extended to massless particles
in~\cite{DelDuca:1990gz}. In QCD, next-to-soft effects have been
investigated using a variety of techniques, including diagrammatic
approaches~\cite{Laenen:2008gt,Laenen:2010uz,Bonocore:2015esa,Bonocore:2016awd,Bonocore:2020xuj,Gervais:2017yxv,
  Gervais:2017zky,Gervais:2017zdb,Laenen:2020nrt,Bahjat-Abbas:2019fqa,vanBeekveld:2021mxn}
that aim to extend the factorisation formula of eq.~(\ref{softcolfac})
to NLP order and / or resum NLP logs, insights from fixed-order
perturbation theory~\cite{Soar:2009yh,
  Moch:2009hr,Moch:2009mu,deFlorian:2014vta,LoPresti:2014ihe,DelDuca:2017twk,vanBeekveld:2019prq,Bonocore:2014wua,Bahjat-Abbas:2018hpv,Ebert:2018lzn,Boughezal:2018mvf,Boughezal:2019ggi},
and effective field
theory~\cite{Kolodrubetz:2016uim,Moult:2016fqy,Feige:2017zci,Beneke:2017ztn,Beneke:2018rbh,Bhattacharya:2018vph,Beneke:2019kgv,Bodwin:2021epw,Moult:2019mog,Beneke:2019oqx,Liu:2019oav,Liu:2020tzd,Boughezal:2016zws,Moult:2017rpl,Chang:2017atu,Moult:2018jjd,Beneke:2018gvs,Ebert:2018gsn,Beneke:2019mua,Moult:2019uhz,Liu:2020ydl,Liu:2020eqe,Wang:2019mym,Beneke:2020ibj}. We
are just starting to learn how to resum NLP effects for certain
specific processes, and the coming years are likely to see a rapid
increasing of our understanding of their importance. Furthermore,
extensions to the factorisation formula of eq.~(\ref{softcolfac})
involve new universal objects in field theory, that may be of interest
beyond QCD, for those wishing to classify new structures in
e.g. ${\cal N}=4$ SYM theory.

We have here seen some basic ideas from threshold resummation, which
involves the inclusion of arbitrary amounts of soft / collinear
radiation. However, the idea of summing up corrections to all orders
in perturbation theory clearly generalises. Another kinematic limit
that has been widely studied is the {\it Regge limit}, in which the
centre of mass energy is much larger than the momentum transfer. This
limit becomes experimentally more relevant as the energy of particle
accelerators increases, and there is significant evidence that the
inclusion of enhanced effects in this limit is needed to better
describe scattering data (see
e.g. refs.~\cite{White:2006yh,Ball:2017otu,Andersen:2020yax,Andersen:2019yzo,Andersen:2018kjg,Andersen:2018tnm,Andersen:2017kfc,Andersen:2016vkp,Andersen:2012gk,Andersen:2011hs,Andersen:2009nu}). This
makes direct contact with much earlier work on S-matrix theory (see
the classic texts of refs.~\cite{Eden:1966dnq,Collins:1977jy}, and
ref.~\cite{White:2019ggo} for a more modern review), ideas from which
(e.g. the bootstrap) seem to be coming back into fashion in
contemporary \texttt{hep-th} physics. Of course, with any resummation,
only a subset of the full information at each order in perturbation
theory is included at higher orders. Resummation thus proceeds in
tandem with fixed-order perturbation theory to provide a two-pronged
attack on QFT: given any observable we want to calculate, we must
include as much information as possible. Low-order information in
perturbation theory can be calculated exactly, and then supplemented
with higher-order results from resummation, being careful that no
contributions have been counted twice. There are then two frontiers in
perturbation theory, namely the inclusion of subleading terms in both
the coupling (NLO, NNLO, ...) and logarithmic (NLL, NNLL, ...)
expansions. Each of these requires clever thinking and new techniques,
and also a joined-up approach. For example, new methods for obtaining
fixed-order results can often be recycled for calculating higher-order
Wilson line correlators, which are needed for resummation. There is
significant scope for more formal amplitudists to contribute to both
areas.

\section{From theory to experiment}
\label{sec:experiment}

\begin{figure}
  \begin{center}
    \scalebox{0.1}{\includegraphics{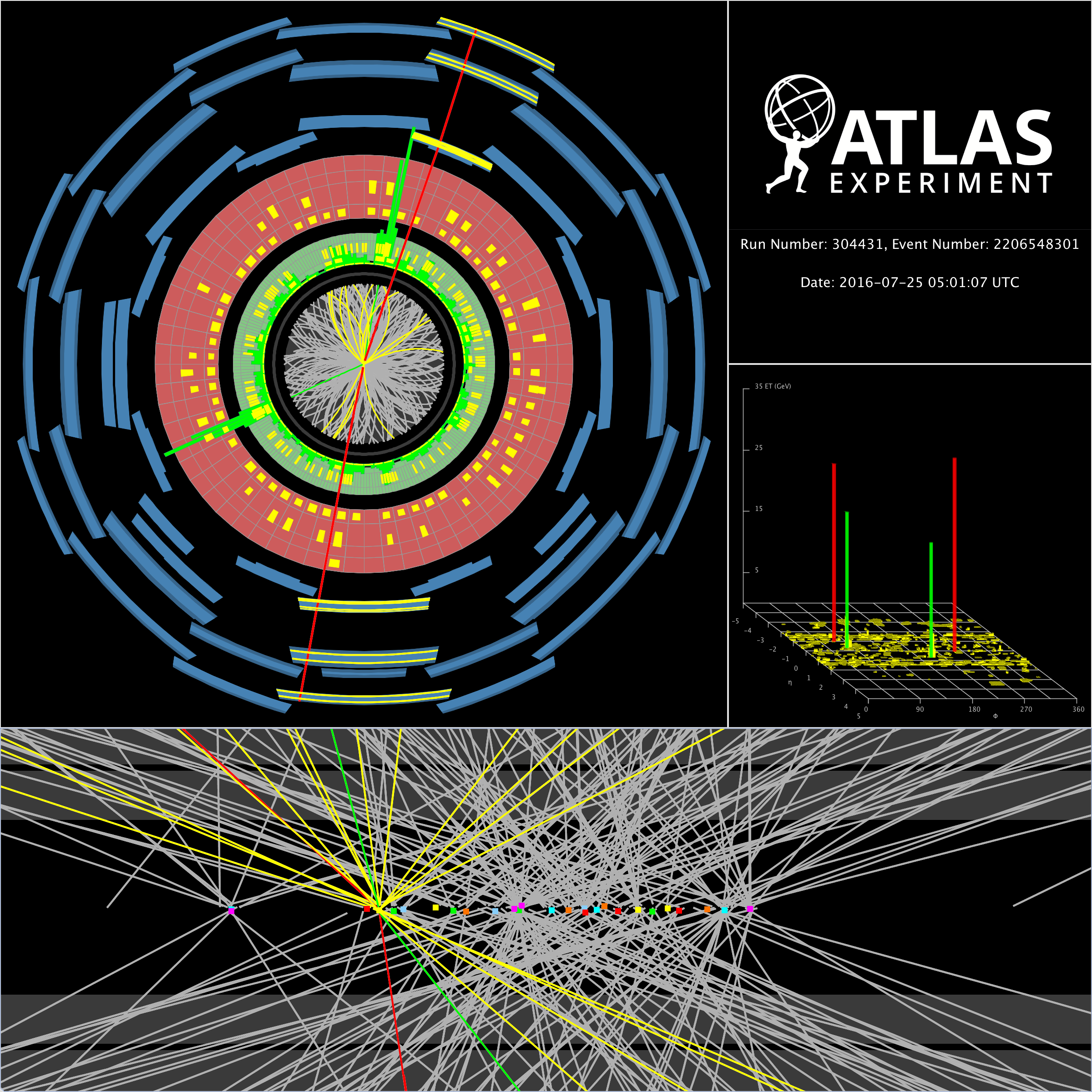}}
    \caption{Event display from the ATLAS experiment, for a candidate
      event in which a Higgs boson decays to two Z bosons. The event
      was used in the analysis of ref.~\cite{ATLAS:2017qey}.}
    \label{fig:atlashzz}
  \end{center}
\end{figure}

In the previous two sections, we have sketched the status of
modern-day efforts to calculate perturbative QFT observables for
collider physics. Anyone working in this area knows how
extraordinarily difficult it can be to squeeze new results out of a
non-abelian gauge theory. However, as we already hinted at in
section~\ref{sec:amplitudes}, the output of even the most intricate
QFT calculation looks almost nothing like what comes out of a particle
accelerator! To illustrate this, figure~\ref{fig:atlashzz} shows an
event that was measured by the ATLAS detector in 2016, and which was
believed to be a Higgs boson decaying to two Z bosons, which
themselves then decay to a pair of leptons each. The upper-left panel
shows a cross-sectional slice through the cylindrical detector, where
the red and green lines constitute the best guess for what the leptons
did. However, there is an enormous number of additional particles. The
yellow lines denote extra charged particles that accompanied the Higgs
boson event. Many of these will be charged hadrons, which arose from
additional quark and gluon radiation. The sheer number of these goes
way beyond what we can reliably calculate in fixed-order perturbation
theory. What's more, only a tiny fraction of these charged particles
have been kept in the event display - those with sufficiently small
momentum transverse to the beam direction are thrown away so that we
can even see what is going on! The grey lines demonstrate an
additional complication: the beams at the LHC do not consist of single
protons, but bunches of many protons (with over 100 billion protons
per bunch in fact). This means that a large number of collisions
happen simultaneously, so that any event we want to look at is swamped
not just by its own mess, but by that of many other independent
collisions! This is called {\it pile-up}, and can be corrected for by
carefully ascertaining that the various particle tracks originate from
different scattering vertices, as is shown in the lower panel of the
figure. However, it is clear that many theorist's idea of what
``comparing theory to data'' means, is very far indeed from what
actually happens in practice. Our aim here is to provide a brief
review on how QFT can nevertheless be meaningfully applied to the
analysis of scattering events, and to point out some of the open
issues.

Theory calculations have two uses. Either we are calculating a signal
(e.g. a new physics process), in which case we want to compare our
prediction for a particular observable (e.g. a cross-section) with
something measured. Or we might be considering a {\it background},
namely a Standard Model process that is in principle different to the
signal, but which may contribute a proportion of events that happen to
look similar. If the background process is much more probable than the
signal, we will get lots of events that mimic the signal and thus act
as ``fake news''. Experimentalists apply very stringent statistical
techniques to make sure that any signal they find has negligible
probability of having been caused by a background. However, to do
this, they need to know the background processes {\it very very
  precisely}, and this is then the job of theorists. But how do the
latter turn their calculations into something approaching what is seen
in the collider?

\begin{figure}
  \begin{center}
    \scalebox{0.6}{\includegraphics{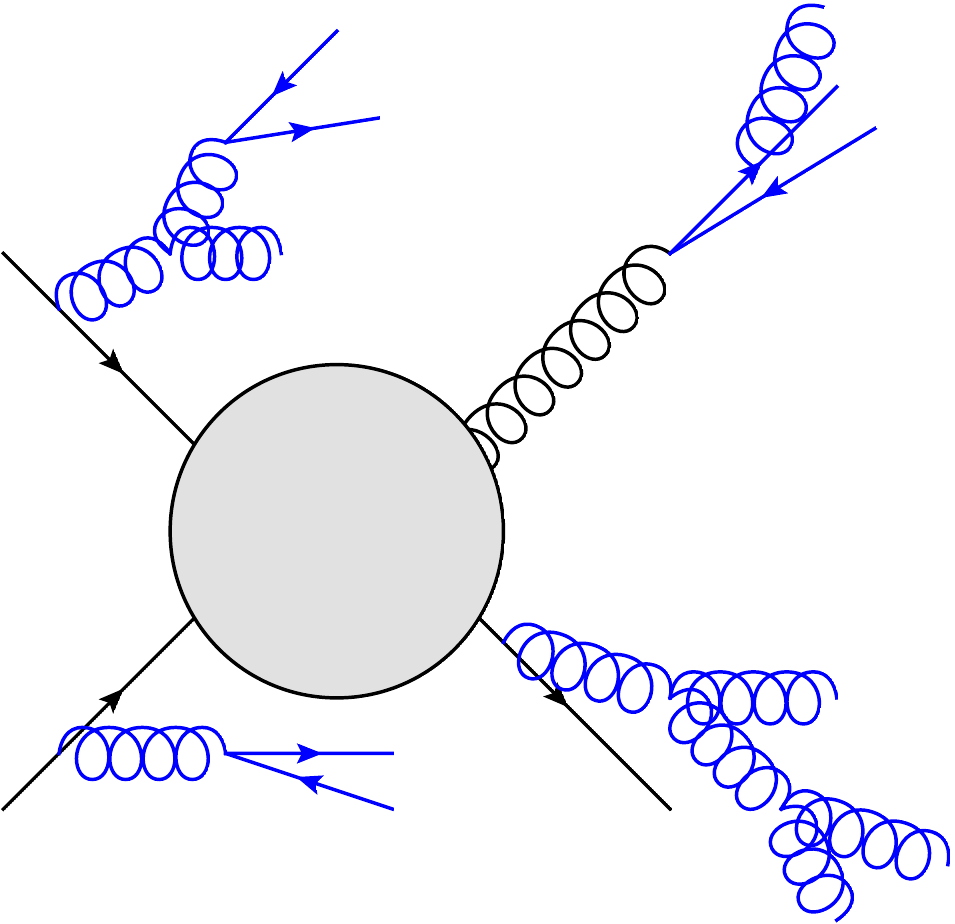}}
    \caption{A {\it parton shower} algorithm dresses a given hard
      scattering process (whose particles are shown here in black)
      with additional quark and gluon radiation (shown in blue). The
      generated probability for such a process is approximate, but
      exact in the limit in which the emitted radiation is collinear
      with the outgoing hard particles. The approximation can be
      improved by including higher-order information in the hard
      scattering process. }
    \label{fig:shower}
  \end{center}
\end{figure}

The first problem is to try to estimate the effect of large amounts of
additional quark and gluon radiation, and there is a very
well-established way to do this known as a {\it parton shower} (see
e.g. refs.~\cite{Campbell:2017hsr,Ellis:1996mzs} for reviews). It
relies on the observation made above, that collinear radiation is
enhanced, and also factorises off from an underlying hard scattering
process in a universal (process-independent) manner. The radiation is
in fact described by known {\it splitting functions} which, roughly
speaking, give the probability that a parton splits into two other
partons, each carrying a certain momentum fraction of the parent. The
factorisation property means that different collinear splittings are
uncorrelated and independent, and this allows for the construction of
an algorithm for the generation of arbitrary amounts of collinear
radiation, modelled as a Markov chain process. Essentially, one can
generate a set of 4-momenta for the additional particles, whose
probability is given by a known distribution, which becomes exact in
the limit in which all the radiation is strictly collinear. It
includes both real and virtual QFT corrections, so that all
probability weights are infrared finite, and applies this distribution
even for radiation that is not collinear. That this turns out to be a
reasonable approximation follows from the fact that collinear
radiation is anyway enhanced. A schematic view of the action of a
parton shower is shown in figure~\ref{fig:shower}.

The energy of modern colliders such as the LHC is such that many
particles are produced, which are widely separated in the detector and
thus not necessarily collinear. Then the parton shower approximation
is insufficient, and must be made more accurate. One way to do this is
to include also full higher-order tree-level amplitudes, which can be
done up to a few extra legs. One can then invent a suitable {\it
  matching prescription}, that guarantees that the most widely
separated particles in any event are described by the tree-level
amplitudes (where these are most accurate), but the remaining
particles with less separation are described by the parton
shower. Several prescriptions exist (see
e.g. refs.~\cite{Caravaglios:1998yr,Catani:2001cc} for some of the
first), and they must carefully make sure that no radiation is
double-counted, by being included in both the higher-order tree-level
matrix elements {\it and} the parton shower. From a QFT point of view,
including only tree-level amplitudes is not ideal, given that the
formal accuracy of the total cross-section then remains at LO
only. Thus, it is desirable to start with full NLO amplitudes, and to
match these with a parton shower. The matching procedure is now even
more delicate than the tree-level case discussed above: the NLO matrix
elements contain both real and virtual corrections, both of which
potentially overlap with what the parton shower is doing. Two NLO
matching schemes are in widespread
use~\cite{Frixione:2002ik,Frixione:2003ei}, and their predictions for
a given process are often compared as a means of estimating the
reliability of the matching procedure. The state-of-the art for many
LHC analyses is that either tree-level amplitudes matched to a parton
shower are used, or the NLO approach (which only includes one
additional emission in the amplitude). The choice of which approach to
use is dictated by what is deemed to be more accurate for the
observable of interest. However, one can clearly go further than this,
by combining as many N$^n$LO matrix elements as possible (including
higher-order tree-level matrix elements), and matching the whole lot
to a parton shower. The subtleties in doing this are many, and the
computational expense of doing this means that more understanding is
needed of where such corrections are genuinely important. This is a
topic that will evolve a great deal in the coming
years~\cite{Re:2021vzu}.

After a parton shower has been applied, our QFT results start to look
a lot more like those in figure~\ref{fig:atlashzz}. But they still
contain free (anti-)quarks and gluons in the final state, rather than
the colour-singlet hadrons that we now are observed in real
experiments, due to the confinement property of QCD. For many
observables this is not a problem. If all we want to do is to estimate
a total cross-section, for example, every parton will end up in {\it
  some} hadron, so the observables calculated assuming final-state
partons, or final-state hadrons must be the same. But if we want to
more realistically model scattering events for use in experimental
analyses, we do indeed want to estimate how the process of {\it
  hadronisation} changes the final state. In practice, this is done
using a variety of phenomenological models, containing free parameters
that can be tuned to data. What helps is that the quantitative effect
of hadronisation on a given observable is known to be suppressed by
powers of the QCD confinement scale $\Lambda$ i.e. it constitutes a
power correction as appears in eq.~(\ref{parton2}).

After parton showering and hadronisation, our scattering events are
still not fully realistic: we have to include the fact that the
incoming partons were only part of the incoming protons. The rest of
the latter will somehow be distributed throughout a given scattering
event, and their colour information may be non-trivially entangled
with the rest of the event. Various models exist for describing this
mess, which is loosely referred to as the {\it underlying event}. New
ideas are always needed, as the uncertainties involved in such models
can have dramatic effects. For example, the mass of the top quark is
currently measured to within about 0.5 GeV, which is very similar to
the estimated theory uncertainty due to underlying event effects (in
particular how the colour of the top quark interacts with the colour
of the beam remnants). This may not sound like a problem, but the top
quark mass enters the expression for quantum corrections to the
potential energy of the Higgs field. This can become negative for a
certain range of top mass values, which results in the vacuum of our
universe becoming unstable. Given the measured Higgs mass value, the
mass of the top quark we currently observe is perilously close to the
unstable range. We can't quite say how close, without knowing the top
quark mass more precisely, and thus our ability to settle our
collective fate rests on better modelling of non-perturbative QCD!

The above steps are computationally technical, and clearly very
complicated for the uninitiated. However, various general purpose
computer programs exist that allow users to simulate scattering events
for a given process, including higher-order amplitudes, parton
showers, hadronisation, underlying event modelling, and more
besides. They are usually referred to as {\it Monte Carlo event
  generators}, and popular programs
include~\cite{Sjostrand:2019zhc,Sherpa:2019gpd,Bellm:2019zci,Alwall:2014hca}. The
output of such programs consists of a set of simulated scattering
events, comprising a list of particles (e.g. leptons, photons,
hadrons) and their 4-momenta. One may then write additional code for
analysing these events, e.g. to select those events that look
interesting, and then plot distributions of various measurable
quantities, to mimic what is done in an actual collider experiment.

Monte Carlo event generators are widely used both by experimentalists,
but also phenomenologists who are trying to find new and better things
to measure. They can also be interfaced with further simulations, that
mimic the behaviour of the real ATLAS and CMS detectors, for an even
more realistic characterisation of how actual scattering events are
likely to behave. This then gives various levels at which theorists
may compare their results with data, and examples include:
\begin{itemize}
\item {\it Parton level}: in this case, experimentalists will try to
  correct for non-perturbative effects, decays of particles etc., and
  present results for (differential) cross-sections containing
  final-state partons, vector bosons, top quarks etc. Theorists can
  then calculate simple final states involving quarks, gluons and
  other particles, using their favourite methods for calculating
  scattering amplitudes, for direct comparison with the ``data''. This
  approach is going out of fashion for differential observables, due
  to the many assumptions that go in to {\it unfolding} the raw data
  back to parton level. If these assumptions turn out to be incorrect
  or superceded years after an experiment finishes, it can be
  impossible to replace them with a more correct analysis, especially
  if the raw data is no longer available.
\item {\it Particle / hadron level}: Here the theory calculation will
  include a parton shower, hadronisation, underlying event etc., as is
  obtained as the output of a Monte Carlo event generator. For some
  very inclusive observables (e.g. total cross-sections), this is not
  necessary, given that the additional steps needed to get to particle
  level will not change the total cross-section. However, the output
  of a particle-level calculation is a set of simulated scattering
  events that looks much closer to what happens in a real detector.
  In particular, this allows theorists to simulate the effect of
  proposed experimental analyses, including how ``interesting''
  scattering events will be selected, and what the distributions of
  various measured quantities will look like.
\item {\it Detector level}: This is similar to particle-level, but
  includes an additional detector simulation, which would be important
  for theorists if they believed that realistic detector effects
  (e.g. finite energy / momentum resolution, gaps in where particles
  can be recorded) may be affecting their predictions.
\end{itemize}
For differential cross-sections, particle level has become a standard
way of comparing theory with data. Even in that case, however, further
steps are usually needed to make experimental events look similar to
particle-level predictions. Both experimental scattering events and
simulated particle-level ones will contain hundreds of particles, most
of which will be hadrons, due to the large amounts of strongly-coupled
quark and gluon radiation. In order to simplify each event, we can
cluster these particles into {\it jets}, using a suitable {\it jet
  algorithm}. We can then select interesting events based on how many
jets they have, and what the distribution of these jets looks like in
the detector. A surprising amount of care is needed to make sure that
a given definition of how to cluster particles into jets is
well-defined in perturbation theory (see ref.~\cite{Salam:2010nqg} for
an excellent review).

The detailed comparison of theory with data requires a constant
dialogue between theorists and experimentalists. At the heart of all
event generators that are used in this process lie our beloved
scattering amplitudes, and thus new techniques from formal theory can
clearly contribute to the ongoing vast international efforts to
understand what the universe is trying to tell us in our detectors.

\section{Summary}
\label{sec:summary}

In the previous sections, we have seen a large number of steps in
between the calculation of scattering amplitudes, and the direct
comparison of theory with quantities measured by experimentalists. It
is thus useful to have a quick summary of these ideas, with some
further context of how they are applied in practice:
\begin{enumerate}
\item[(a)] Squared amplitudes must be renormalised to remove UV singularities, and
  converted into cross-sections by integrating over the phase space of
  any final-state particles, and dividing by the Lorentz-invariant
  flux factor.
\item At hadron colliders, one must combine cross-sections for
  incoming {\it partons} (i.e. (anti-)quarks and gluons) with {\it
    parton distribution} functions for the incoming beam particles
  (e.g. protons).
\item[(b)] Theoretical quantities (e.g. (differential) cross-sections)
  should be such that they are {\it infrared-safe}: final states with
  different numbers of particles must be combined so that there are no
  IR singularities in the final results. One should then include as
  many perturbative orders in each amplitude as possible.
\item[(c)] If the coefficients of the perturbation expansion are unstable,
  it may be necessary to include additional logarithmically-enhanced
  terms to all orders in perturbation theory (``resummation''), and to
  avoid double-counting contributions which have already been included
  in the fixed-order expansion.
\item[(d)] One can approximate the emission of additional QCD radiation by
  applying a {\it parton shower} algorithm to the theoretical
  prediction for a given final state. Care is needed if combining
  higher-order amplitudes with the parton shower, to avoid
  double-counting contributions.
\item[(e)] After the parton shower, one may include further algorithms that
  simulate the combination of outgoing partons into colour-singlet
  hadrons, interactions with the beam remnants etc.
\end{enumerate}
Depending on what data we are comparing to, it is not necessary to
include all of these steps. To clarify this, let us consider one of
the simplest types of observable that an experiment might measure,
namely the cross-section for a given process. Experiments sometimes
present a best estimate for the total cross-section, having attempted
to reconstruct any intermediate unstable particles, and corrected for
the finite detector volume etc. This is perhaps the easiest type of
data for a theorist to compare to. If they wish, they can use
fixed-order perturbation theory to obtain the total cross-section
(including the parton distributions if it is a hadron collider) at
some order, and then directly compare this number with the data. The
theory result will have some uncertainty estimated by varying the
renormalisation and factorisation scales. Likewise, the experimental
result will also have an uncertainty, which may be split further into
estimated {\it systematic} and {\it statistical errors}. The first of
these is due to possible biases in the measurement, whereas the second
measures the limited power of a finite data set. If the theory and
data numbers agree within their respective uncertainties, then we
would say that the theory matches the data. If not, we can
statistically quantify the level of disagreement. This corresponds to
the {\it parton level} comparison outlined above.

Even for total cross-sections, there are more sophisticated
approaches. For example, we may wish to resum contributions from
certain kinematic regions, the results of which are typically
available in dedicated public codes for given processes. It is
routine, for example, to include resummed contributions for various
top quark and Higgs boson cross-sections. The effect of including
these resummed contributions may be to change the central value of our
theory prediction, or modify its uncertainty. However, it is still a
single number which we can compare with the measured result. An even
more complicated approach would be to dress our theoretical prediction
with a parton shower algorithm, hadronisation etc. This simulates
final states with large numbers of particles, and we can then apply
similar selection criteria to those used in the experiment to our
simulated events, in order to reproduce the steps they used in
performing their own measurement. This is particularly useful if the
experiment presents a measured cross-section that does not correspond
to the total cross-section one would calculate in QFT, but has some
exclusion criteria already applied (e.g. requirements that the
particles be in certain regions of the detector only).

Other common measurements are differential cross-sections involving
numbers of particles or jets, or their kinematic properties
(e.g. angles, transverse momenta etc.). For low numbers of particles,
it may be possible to rely purely on fixed-order perturbation theory,
provided there are enough particles in the final state of the relevant
amplitudes to coincide with all the particles one wants for a given
kinematic distribution. However, a much more realistic estimate will
be obtained by using parton-shower programs to simulate realistic
events containing many particles, and then clustering the particles
into jets using the same algorithms that the experiment has used. In
order to do this, a theorist would use one of the publicly available
Monte Carlo event generators mentioned above.

Concrete examples of processes at the LHC for which the above remarks
apply are perhaps too numerous to list explicitly. Furthermore,
modern-day Monte Carlo event generators essentially allow the user to
specify {\it any} desired process, and then to automatically generate
simulated events which can be analysed to produce kinematic
distributions for comparison with real data. A particularly convenient
repository for experimental data in particle physics is the HEPData
database~\cite{HEPData}, which also provides links to the relevant
publications.

Finally, it is worth commenting on what theory input is missing for
forthcoming data analyses and / or new collider experiments beyond the
currently running LHC. Broadly speaking, the answer is that we must
learn to compute ever-higher orders in perturbation theory, and to be
able to match amplitudes calculated at these orders to
state-of-the-art parton-shower algorithms, where the latter can also
be improved. Given the large amount of effort involved, it is
necessary to prioritise those processes that experiments are
particularly focused on. To this end, wishlists such as those in
ref.~\cite{Huss:2022ful} are highly useful, as they distil the opinions
of large numbers of experimentalists into a coherent request!  Another
area requiring improvement is the determination of the parton
distribution functions, as the uncertainty on extracting these from
data can limit the theoretical precision of a given collider
prediction, if the relevant cross-section is known to a very high
perturbative order. It is currently not definitively known what the
next collider after the LHC will be. However, it is highly likely to
be a lepton-based machine (e.g. colliding electrons and
positrons). Many high-precision cross-sections are already known for
such colliders, given that the calculations are actually somewhat
simpler than those needed for hadron machines, due to the lack of
coloured particles in the initial state. But it is always useful to
have higher orders in perturbation theory, as for hadron colliders.

\section{Conclusion}
\label{sec:conclude}

It is well-known that scattering amplitudes in perturbative QFT are
relevant for collider experiments. But the precise way in which they
are used, including the huge number of steps that are needed to
meaningfully compare theory with data, are never dealt with in
textbooks. It is perfectly possible nowadays for formal theorists to
live out an entire career without comparing anything to data, despite
a desire to do so. It is also very often true that the topics being
discussed on \texttt{hep-th} (e.g. Wilson lines, (next-to)-soft
radiation) are the {\it same} topics being discussed on
\texttt{hep-ph}, albeit in completely different notation, and for
entirely different reasons. Hence, there is significant scope for more
formal theorists to contribute to the challenging and rewarding world
of collider physics, should they wish to do so. Recent years have seen
remarkable revolutions in how we think about scattering amplitudes in
different theories, and how those different theories are related. My
hope -- which is also the very ethos of the SAGEX network -- is that
the coming years will see similar revolutions in how we apply new
techniques to extract the most we can from current and forthcoming
collider experiments.

\section*{Acknowledgments}
This work was supported by the European Union's Horizon 2020 research
and innovation programme under the Marie Sk\l{}odowska-Curie grant
agreement No.~764850 {\it ``\href{https://sagex.org}{SAGEX}''}. We
also acknowledge support from the Science and Technology Facilities
Council (STFC) Consolidated Grant ST/T000686/1 \textit{``Amplitudes,
  strings \& duality''}.\\

\bibliography{white.bib}

\providecommand{\newblock}{}
\begin{thebibliography}{100}
\expandafter\ifx\csname url\endcsname\relax
  \def\url#1{{\tt #1}}\fi
\expandafter\ifx\csname urlprefix\endcsname\relax\def\urlprefix{URL }\fi
\providecommand{\eprint}[2][]{\url{#2}}

\bibitem{ATLAS:2019zrq}
Aaboud M {\em et~al.\/} (ATLAS) 2020 {\em Eur. Phys. J. C\/} {\bf 80} 754
  (\textit{Preprint} \eprint{1903.07570})

\bibitem{Anastasiou:2015vya}
Anastasiou C, Duhr C, Dulat F, Herzog F and Mistlberger B 2015 {\em Phys. Rev.
  Lett.\/} {\bf 114} 212001 (\textit{Preprint} \eprint{1503.06056})

\bibitem{Anastasiou:2016cez}
Anastasiou C, Duhr C, Dulat F, Furlan E, Gehrmann T, Herzog F, Lazopoulos A and
  Mistlberger B 2016 {\em JHEP\/} {\bf 05} 058 (\textit{Preprint}
  \eprint{1602.00695})

\bibitem{Dreyer:2016oyx}
Dreyer F~A and Karlberg A 2016 {\em Phys. Rev. Lett.\/} {\bf 117} 072001
  (\textit{Preprint} \eprint{1606.00840})

\bibitem{Dreyer:2018qbw}
Dreyer F~A and Karlberg A 2018 {\em Phys. Rev. D\/} {\bf 98} 114016
  (\textit{Preprint} \eprint{1811.07906})

\bibitem{Duhr:2020seh}
Duhr C, Dulat F and Mistlberger B 2020 {\em Phys. Rev. Lett.\/} {\bf 125}
  172001 (\textit{Preprint} \eprint{2001.07717})

\bibitem{Gross:1973id}
Gross D~J and Wilczek F 1973 {\em Phys. Rev. Lett.\/} {\bf 30} 1343--1346

\bibitem{Politzer:1973fx}
Politzer H~D 1973 {\em Phys. Rev. Lett.\/} {\bf 30} 1346--1349

\bibitem{Feynman:1969wa}
Feynman R~P 1969 {\em Conf. Proc. C\/} {\bf 690905} 237--258

\bibitem{Feynman:1969ej}
Feynman R~P 1969 {\em Phys. Rev. Lett.\/} {\bf 23} 1415--1417

\bibitem{Collins:1984xc}
Collins J~C 1986 {\em {Renormalization}: {An Introduction to Renormalization,
  The Renormalization Group, and the Operator Product Expansion}\/} ({\em
  Cambridge Monographs on Mathematical Physics\/} vol~26) (Cambridge: Cambridge
  University Press) ISBN 978-0-521-31177-9, 978-0-511-86739-2

\bibitem{Bloch:1937pw}
Bloch F and Nordsieck A 1937 {\em Phys. Rev.\/} {\bf 52} 54--59

\bibitem{Kinoshita:1962ur}
Kinoshita T 1962 {\em J. Math. Phys.\/} {\bf 3} 650--677

\bibitem{Lee:1964is}
Lee T~D and Nauenberg M 1964 {\em Phys. Rev.\/} {\bf 133} B1549--B1562

\bibitem{Bailey:2020ooq}
Bailey S, Cridge T, Harland-Lang L~A, Martin A~D and Thorne R~S 2021 {\em Eur.
  Phys. J. C\/} {\bf 81} 341 (\textit{Preprint} \eprint{2012.04684})

\bibitem{NNPDF:2019ubu}
Abdul~Khalek R {\em et~al.\/} (NNPDF) 2019 {\em Eur. Phys. J. C\/} {\bf 79} 931
  (\textit{Preprint} \eprint{1906.10698})

\bibitem{Alekhin:2018pai}
Alekhin S, Bl\"umlein J and Moch S 2018 {\em Eur. Phys. J. C\/} {\bf 78} 477
  (\textit{Preprint} \eprint{1803.07537})

\bibitem{Hou:2019efy}
Hou T~J {\em et~al.\/} 2021 {\em Phys. Rev. D\/} {\bf 103} 014013
  (\textit{Preprint} \eprint{1912.10053})

\bibitem{Passarino:1978jh}
Passarino G and Veltman M~J~G 1979 {\em Nucl. Phys. B\/} {\bf 160} 151--207

\bibitem{Ellis:2007qk}
Ellis R~K and Zanderighi G 2008 {\em JHEP\/} {\bf 02} 002 (\textit{Preprint}
  \eprint{0712.1851})

\bibitem{Bern:1994zx}
Bern Z, Dixon L~J, Dunbar D~C and Kosower D~A 1994 {\em Nucl. Phys. B\/} {\bf
  425} 217--260 (\textit{Preprint} \eprint{hep-ph/9403226})

\bibitem{Bern:1994cg}
Bern Z, Dixon L~J, Dunbar D~C and Kosower D~A 1995 {\em Nucl. Phys. B\/} {\bf
  435} 59--101 (\textit{Preprint} \eprint{hep-ph/9409265})

\bibitem{Bern:1995db}
Bern Z and Morgan A~G 1996 {\em Nucl. Phys. B\/} {\bf 467} 479--509
  (\textit{Preprint} \eprint{hep-ph/9511336})

\bibitem{Bern:1997sc}
Bern Z, Dixon L~J and Kosower D~A 1998 {\em Nucl. Phys. B\/} {\bf 513} 3--86
  (\textit{Preprint} \eprint{hep-ph/9708239})

\bibitem{Britto:2004nc}
Britto R, Cachazo F and Feng B 2005 {\em Nucl. Phys. B\/} {\bf 725} 275--305
  (\textit{Preprint} \eprint{hep-th/0412103})

\bibitem{Brandhuber:2005jw}
Brandhuber A, McNamara S, Spence B~J and Travaglini G 2005 {\em JHEP\/} {\bf
  10} 011 (\textit{Preprint} \eprint{hep-th/0506068})

\bibitem{Britto:2004ap}
Britto R, Cachazo F and Feng B 2005 {\em Nucl. Phys. B\/} {\bf 715} 499--522
  (\textit{Preprint} \eprint{hep-th/0412308})

\bibitem{Bern:2005hs}
Bern Z, Dixon L~J and Kosower D~A 2005 {\em Phys. Rev. D\/} {\bf 71} 105013
  (\textit{Preprint} \eprint{hep-th/0501240})

\bibitem{Bern:2005ji}
Bern Z, Dixon L~J and Kosower D~A 2005 {\em Phys. Rev. D\/} {\bf 72} 125003
  (\textit{Preprint} \eprint{hep-ph/0505055})

\bibitem{Badger:2005zh}
Badger S~D, Glover E~W~N, Khoze V~V and Svrcek P 2005 {\em JHEP\/} {\bf 07} 025
  (\textit{Preprint} \eprint{hep-th/0504159})

\bibitem{Badger:2005jv}
Badger S~D, Glover E~W~N and Khoze V~V 2006 {\em JHEP\/} {\bf 01} 066
  (\textit{Preprint} \eprint{hep-th/0507161})

\bibitem{Anastasiou:2006jv}
Anastasiou C, Britto R, Feng B, Kunszt Z and Mastrolia P 2007 {\em Phys. Lett.
  B\/} {\bf 645} 213--216 (\textit{Preprint} \eprint{hep-ph/0609191})

\bibitem{Anastasiou:2006gt}
Anastasiou C, Britto R, Feng B, Kunszt Z and Mastrolia P 2007 {\em JHEP\/} {\bf
  03} 111 (\textit{Preprint} \eprint{hep-ph/0612277})

\bibitem{Giele:2008ve}
Giele W~T, Kunszt Z and Melnikov K 2008 {\em JHEP\/} {\bf 04} 049
  (\textit{Preprint} \eprint{0801.2237})

\bibitem{Huang:2013vha}
Huang Y~t and McGady D 2014 {\em Phys. Rev. Lett.\/} {\bf 112} 241601
  (\textit{Preprint} \eprint{1307.4065})

\bibitem{Ellis:2011cr}
Ellis R~K, Kunszt Z, Melnikov K and Zanderighi G 2012 {\em Phys. Rept.\/} {\bf
  518} 141--250 (\textit{Preprint} \eprint{1105.4319})

\bibitem{Gleisberg:2007md}
Gleisberg T and Krauss F 2008 {\em Eur. Phys. J. C\/} {\bf 53} 501--523
  (\textit{Preprint} \eprint{0709.2881})

\bibitem{Hirschi:2011pa}
Hirschi V, Frederix R, Frixione S, Garzelli M~V, Maltoni F and Pittau R 2011
  {\em JHEP\/} {\bf 05} 044 (\textit{Preprint} \eprint{1103.0621})

\bibitem{Bellm:2015jjp}
Bellm J {\em et~al.\/} 2016 {\em Eur. Phys. J. C\/} {\bf 76} 196
  (\textit{Preprint} \eprint{1512.01178})

\bibitem{Sherpa:2019gpd}
Bothmann E {\em et~al.\/} (Sherpa) 2019 {\em SciPost Phys.\/} {\bf 7} 034
  (\textit{Preprint} \eprint{1905.09127})

\bibitem{Alwall:2014hca}
Alwall J, Frederix R, Frixione S, Hirschi V, Maltoni F, Mattelaer O, Shao H~S,
  Stelzer T, Torrielli P and Zaro M 2014 {\em JHEP\/} {\bf 07} 079
  (\textit{Preprint} \eprint{1405.0301})

\bibitem{Bellm:2019zci}
Bellm J {\em et~al.\/} 2020 {\em Eur. Phys. J. C\/} {\bf 80} 452
  (\textit{Preprint} \eprint{1912.06509})

\bibitem{Abreu:2022mfk}
Abreu S, Britto R and Duhr C 2022 {\em J. Phys. A\/} {\bf 55} 443004
  (\textit{Preprint} \eprint{2203.13014})

\bibitem{Blumlein:2022zkr}
Bl\"umlein J and Schneider C 2022 {\em J. Phys. A\/} {\bf 55} 443005
  (\textit{Preprint} \eprint{2203.13015})

\bibitem{Papathanasiou:2022lan}
Papathanasiou G 2022 {\em J. Phys. A\/} {\bf 55} 443006 (\textit{Preprint}
  \eprint{2203.13016})

\bibitem{Travaglini:2022uwo}
Travaglini G {\em et~al.\/} 2022 {\em J. Phys. A\/} {\bf 55} 443001
  (\textit{Preprint} \eprint{2203.13011})

\bibitem{Frixione:1995ms}
Frixione S, Kunszt Z and Signer A 1996 {\em Nucl. Phys. B\/} {\bf 467} 399--442
  (\textit{Preprint} \eprint{hep-ph/9512328})

\bibitem{Catani:1996vz}
Catani S and Seymour M~H 1997 {\em Nucl. Phys. B\/} {\bf 485} 291--419
  [Erratum: Nucl.Phys.B 510, 503--504 (1998)] (\textit{Preprint}
  \eprint{hep-ph/9605323})

\bibitem{Gehrmann-DeRidder:2005btv}
Gehrmann-De~Ridder A, Gehrmann T and Glover E~W~N 2005 {\em JHEP\/} {\bf 09}
  056 (\textit{Preprint} \eprint{hep-ph/0505111})

\bibitem{Currie:2013vh}
Currie J, Glover E~W~N and Wells S 2013 {\em JHEP\/} {\bf 04} 066
  (\textit{Preprint} \eprint{1301.4693})

\bibitem{Czakon:2010td}
Czakon M 2010 {\em Phys. Lett. B\/} {\bf 693} 259--268 (\textit{Preprint}
  \eprint{1005.0274})

\bibitem{Czakon:2011ve}
Czakon M 2011 {\em Nucl. Phys. B\/} {\bf 849} 250--295 (\textit{Preprint}
  \eprint{1101.0642})

\bibitem{Boughezal:2011jf}
Boughezal R, Melnikov K and Petriello F 2012 {\em Phys. Rev. D\/} {\bf 85}
  034025 (\textit{Preprint} \eprint{1111.7041})

\bibitem{Anastasiou:2003gr}
Anastasiou C, Melnikov K and Petriello F 2004 {\em Phys. Rev. D\/} {\bf 69}
  076010 (\textit{Preprint} \eprint{hep-ph/0311311})

\bibitem{Binoth:2004jv}
Binoth T and Heinrich G 2004 {\em Nucl. Phys. B\/} {\bf 693} 134--148
  (\textit{Preprint} \eprint{hep-ph/0402265})

\bibitem{Catani:2007vq}
Catani S and Grazzini M 2007 {\em Phys. Rev. Lett.\/} {\bf 98} 222002
  (\textit{Preprint} \eprint{hep-ph/0703012})

\bibitem{Boughezal:2015eha}
Boughezal R, Liu X and Petriello F 2015 {\em Phys. Rev. D\/} {\bf 91} 094035
  (\textit{Preprint} \eprint{1504.02540})

\bibitem{Gaunt:2015pea}
Gaunt J, Stahlhofen M, Tackmann F~J and Walsh J~R 2015 {\em JHEP\/} {\bf 09}
  058 (\textit{Preprint} \eprint{1505.04794})

\bibitem{DelDuca:2015zqa}
Del~Duca V, Duhr C, Somogyi G, Tramontano F and Tr\'ocs\'anyi Z 2015 {\em
  JHEP\/} {\bf 04} 036 (\textit{Preprint} \eprint{1501.07226})

\bibitem{Caola:2017dug}
Caola F, Melnikov K and R\"ontsch R 2017 {\em Eur. Phys. J. C\/} {\bf 77} 248
  (\textit{Preprint} \eprint{1702.01352})

\bibitem{Delto:2019asp}
Delto M and Melnikov K 2019 {\em JHEP\/} {\bf 05} 148 (\textit{Preprint}
  \eprint{1901.05213})

\bibitem{Caola:2019nzf}
Caola F, Melnikov K and R\"ontsch R 2019 {\em Eur. Phys. J. C\/} {\bf 79} 386
  (\textit{Preprint} \eprint{1902.02081})

\bibitem{Magnea:2018hab}
Magnea L, Maina E, Pelliccioli G, Signorile-Signorile C, Torrielli P and
  Uccirati S 2018 {\em JHEP\/} {\bf 12} 107 [Erratum: JHEP 06, 013 (2019)]
  (\textit{Preprint} \eprint{1806.09570})

\bibitem{Magnea:2018ebr}
Magnea L, Maina E, Pelliccioli G, Signorile-Signorile C, Torrielli P and
  Uccirati S 2018 {\em JHEP\/} {\bf 12} 062 (\textit{Preprint}
  \eprint{1809.05444})

\bibitem{Caola:2019pfz}
Caola F, Melnikov K and R\"ontsch R 2019 {\em Eur. Phys. J. C\/} {\bf 79} 1013
  (\textit{Preprint} \eprint{1907.05398})

\bibitem{Asteriadis:2019dte}
Asteriadis K, Caola F, Melnikov K and R\"ontsch R 2020 {\em Eur. Phys. J. C\/}
  {\bf 80} 8 (\textit{Preprint} \eprint{1910.13761})

\bibitem{Cacciari:2015jma}
Cacciari M, Dreyer F~A, Karlberg A, Salam G~P and Zanderighi G 2015 {\em Phys.
  Rev. Lett.\/} {\bf 115} 082002 [Erratum: Phys.Rev.Lett. 120, 139901 (2018)]
  (\textit{Preprint} \eprint{1506.02660})

\bibitem{Huss:2022ful}
Huss A, Huston J, Jones S and Pellen M 2022  (\textit{Preprint}
  \eprint{2207.02122})

\bibitem{Denner:2019vbn}
Denner A and Dittmaier S 2020 {\em Phys. Rept.\/} {\bf 864} 1--163
  (\textit{Preprint} \eprint{1912.06823})

\bibitem{Agarwal:2021ais}
Agarwal N, Magnea L, Signorile-Signorile C and Tripathi A 2021
  (\textit{Preprint} \eprint{2112.07099})

\bibitem{Parisi:1979xd}
Parisi G 1980 {\em Phys. Lett. B\/} {\bf 90} 295--296

\bibitem{Curci:1979am}
Curci G and Greco M 1980 {\em Phys. Lett. B\/} {\bf 92} 175--178

\bibitem{Sterman:1986aj}
Sterman G~F 1987 {\em Nucl. Phys. B\/} {\bf 281} 310--364

\bibitem{Catani:1989ne}
Catani S and Trentadue L 1989 {\em Nucl. Phys. B\/} {\bf 327} 323--352

\bibitem{Gatheral:1983cz}
Gatheral J~G~M 1983 {\em Phys. Lett. B\/} {\bf 133} 90--94

\bibitem{Frenkel:1984pz}
Frenkel J and Taylor J~C 1984 {\em Nucl. Phys. B\/} {\bf 246} 231--245

\bibitem{Sterman:1981jc}
Sterman G~F 1981 {\em AIP Conf. Proc.\/} {\bf 74} 22--40

\bibitem{Korchemsky:1992xv}
Korchemsky G~P and Marchesini G 1993 {\em Nucl. Phys. B\/} {\bf 406} 225--258
  (\textit{Preprint} \eprint{hep-ph/9210281})

\bibitem{Korchemsky:1993uz}
Korchemsky G~P and Marchesini G 1993 {\em Phys. Lett. B\/} {\bf 313} 433--440

\bibitem{Forte:2002ni}
Forte S and Ridolfi G 2003 {\em Nucl. Phys. B\/} {\bf 650} 229--270
  (\textit{Preprint} \eprint{hep-ph/0209154})

\bibitem{Becher:2006nr}
Becher T and Neubert M 2006 {\em Phys. Rev. Lett.\/} {\bf 97} 082001
  (\textit{Preprint} \eprint{hep-ph/0605050})

\bibitem{Schwartz:2007ib}
Schwartz M~D 2008 {\em Phys. Rev. D\/} {\bf 77} 014026 (\textit{Preprint}
  \eprint{0709.2709})

\bibitem{Bauer:2008dt}
Bauer C~W, Fleming S~P, Lee C and Sterman G~F 2008 {\em Phys. Rev. D\/} {\bf
  78} 034027 (\textit{Preprint} \eprint{0801.4569})

\bibitem{Chiu:2009mg}
Chiu J~y, Fuhrer A, Kelley R and Manohar A~V 2009 {\em Phys. Rev. D\/} {\bf 80}
  094013 (\textit{Preprint} \eprint{0909.0012})

\bibitem{Luisoni:2015xha}
Luisoni G and Marzani S 2015 {\em J. Phys. G\/} {\bf 42} 103101
  (\textit{Preprint} \eprint{1505.04084})

\bibitem{Becher:2014oda}
Becher T, Broggio A and Ferroglia A 2015 {\em {Introduction to Soft-Collinear
  Effective Theory}\/} vol 896 (Springer) (\textit{Preprint}
  \eprint{1410.1892})

\bibitem{Campbell:2017hsr}
Campbell J, Huston J and Krauss F 2017 {\em {The Black Book of Quantum
  Chromodynamics}: {A Primer for the LHC Era}\/} (Oxford University Press) ISBN
  978-0-19-965274-7

\bibitem{Dixon:2008gr}
Dixon L~J, Magnea L and Sterman G~F 2008 {\em JHEP\/} {\bf 08} 022
  (\textit{Preprint} \eprint{0805.3515})

\bibitem{Weinberg:1965nx}
Weinberg S 1965 {\em Phys. Rev.\/} {\bf 140} B516--B524

\bibitem{Alday:2007hr}
Alday L~F and Maldacena J~M 2007 {\em JHEP\/} {\bf 06} 064 (\textit{Preprint}
  \eprint{0705.0303})

\bibitem{Polyakov:1980ca}
Polyakov A~M 1980 {\em Nucl. Phys. B\/} {\bf 164} 171--188

\bibitem{Arefeva:1980zd}
Arefeva I~Y 1980 {\em Phys. Lett. B\/} {\bf 93} 347--353

\bibitem{Dotsenko:1979wb}
Dotsenko V~S and Vergeles S~N 1980 {\em Nucl. Phys. B\/} {\bf 169} 527--546

\bibitem{Brandt:1981kf}
Brandt R~A, Neri F and Sato M~a 1981 {\em Phys. Rev. D\/} {\bf 24} 879

\bibitem{Korchemsky:1985xj}
Korchemsky G~P and Radyushkin A~V 1986 {\em Phys. Lett. B\/} {\bf 171} 459--467

\bibitem{Korchemsky:1985xu}
Korchemsky G~P and Radyushkin A~V 1986 {\em Sov. J. Nucl. Phys.\/} {\bf 44} 877

\bibitem{Korchemsky:1987wg}
Korchemsky G~P and Radyushkin A~V 1987 {\em Nucl. Phys. B\/} {\bf 283} 342--364

\bibitem{Grozin:2015kna}
Grozin A, Henn J~M, Korchemsky G~P and Marquard P 2016 {\em JHEP\/} {\bf 01}
  140 (\textit{Preprint} \eprint{1510.07803})

\bibitem{Grozin:2014hna}
Grozin A, Henn J~M, Korchemsky G~P and Marquard P 2015 {\em Phys. Rev. Lett.\/}
  {\bf 114} 062006 (\textit{Preprint} \eprint{1409.0023})

\bibitem{Bruser:2020bsh}
Br\"user R, Dlapa C, Henn J~M and Yan K 2021 {\em Phys. Rev. Lett.\/} {\bf 126}
  021601 (\textit{Preprint} \eprint{2007.04851})

\bibitem{Korchemsky:1993hr}
Korchemsky G~P 1994 {\em Phys. Lett. B\/} {\bf 325} 459--466 (\textit{Preprint}
  \eprint{hep-ph/9311294})

\bibitem{Korchemskaya:1996je}
Korchemskaya I~A and Korchemsky G~P 1996 {\em Phys. Lett. B\/} {\bf 387}
  346--354 (\textit{Preprint} \eprint{hep-ph/9607229})

\bibitem{Korchemskaya:1994qp}
Korchemskaya I~A and Korchemsky G~P 1995 {\em Nucl. Phys. B\/} {\bf 437}
  127--162 (\textit{Preprint} \eprint{hep-ph/9409446})

\bibitem{Catani:1998bh}
Catani S 1998 {\em Phys. Lett. B\/} {\bf 427} 161--171 (\textit{Preprint}
  \eprint{hep-ph/9802439})

\bibitem{Sterman:2002qn}
Sterman G~F and Tejeda-Yeomans M~E 2003 {\em Phys. Lett. B\/} {\bf 552} 48--56
  (\textit{Preprint} \eprint{hep-ph/0210130})

\bibitem{Kidonakis:1998nf}
Kidonakis N, Oderda G and Sterman G~F 1998 {\em Nucl. Phys. B\/} {\bf 531}
  365--402 (\textit{Preprint} \eprint{hep-ph/9803241})

\bibitem{Bonciani:2003nt}
Bonciani R, Catani S, Mangano M~L and Nason P 2003 {\em Phys. Lett. B\/} {\bf
  575} 268--278 (\textit{Preprint} \eprint{hep-ph/0307035})

\bibitem{Dokshitzer:2005ig}
Dokshitzer Y~L and Marchesini G 2006 {\em JHEP\/} {\bf 01} 007
  (\textit{Preprint} \eprint{hep-ph/0509078})

\bibitem{Aybat:2006mz}
Aybat S~M, Dixon L~J and Sterman G~F 2006 {\em Phys. Rev. D\/} {\bf 74} 074004
  (\textit{Preprint} \eprint{hep-ph/0607309})

\bibitem{Gardi:2009qi}
Gardi E and Magnea L 2009 {\em JHEP\/} {\bf 03} 079 (\textit{Preprint}
  \eprint{0901.1091})

\bibitem{Becher:2009cu}
Becher T and Neubert M 2009 {\em Phys. Rev. Lett.\/} {\bf 102} 162001 [Erratum:
  Phys.Rev.Lett. 111, 199905 (2013)] (\textit{Preprint} \eprint{0901.0722})

\bibitem{Becher:2009qa}
Becher T and Neubert M 2009 {\em JHEP\/} {\bf 06} 081 [Erratum: JHEP 11, 024
  (2013)] (\textit{Preprint} \eprint{0903.1126})

\bibitem{Gardi:2009zv}
Gardi E and Magnea L 2009 {\em Nuovo Cim. C\/} {\bf 32N5-6} 137--157
  (\textit{Preprint} \eprint{0908.3273})

\bibitem{Dixon:2009gx}
Dixon L~J 2009 {\em Phys. Rev. D\/} {\bf 79} 091501 (\textit{Preprint}
  \eprint{0901.3414})

\bibitem{Dixon:2009ur}
Dixon L~J, Gardi E and Magnea L 2010 {\em JHEP\/} {\bf 02} 081
  (\textit{Preprint} \eprint{0910.3653})

\bibitem{DelDuca:2011ae}
Del~Duca V, Duhr C, Gardi E, Magnea L and White C~D 2011 {\em JHEP\/} {\bf 12}
  021 (\textit{Preprint} \eprint{1109.3581})

\bibitem{Caron-Huot:2013fea}
Caron-Huot S 2015 {\em JHEP\/} {\bf 05} 093 (\textit{Preprint}
  \eprint{1309.6521})

\bibitem{Ahrens:2012qz}
Ahrens V, Neubert M and Vernazza L 2012 {\em JHEP\/} {\bf 09} 138
  (\textit{Preprint} \eprint{1208.4847})

\bibitem{Naculich:2013xa}
Naculich S~G, Nastase H and Schnitzer H~J 2013 {\em JHEP\/} {\bf 04} 114
  (\textit{Preprint} \eprint{1301.2234})

\bibitem{Erdogan:2014gha}
Erdo\u{g}an O and Sterman G 2015 {\em Phys. Rev. D\/} {\bf 91} 065033
  (\textit{Preprint} \eprint{1411.4588})

\bibitem{Gehrmann:2010ue}
Gehrmann T, Glover E~W~N, Huber T, Ikizlerli N and Studerus C 2010 {\em JHEP\/}
  {\bf 06} 094 (\textit{Preprint} \eprint{1004.3653})

\bibitem{Agarwal:2021zft}
Agarwal B, von Manteuffel A, Panzer E and Schabinger R~M 2021 {\em Phys. Lett.
  B\/} {\bf 820} 136503 (\textit{Preprint} \eprint{2102.09725})

\bibitem{Almelid:2015jia}
Almelid O, Duhr C and Gardi E 2016 {\em Phys. Rev. Lett.\/} {\bf 117} 172002
  (\textit{Preprint} \eprint{1507.00047})

\bibitem{Gardi:2016ttq}
Gardi E, Almelid O and Duhr C 2016 {\em PoS\/} {\bf LL2016} 058
  (\textit{Preprint} \eprint{1606.05697})

\bibitem{Kidonakis:2009ev}
Kidonakis N 2009 {\em Phys. Rev. Lett.\/} {\bf 102} 232003 (\textit{Preprint}
  \eprint{0903.2561})

\bibitem{Mitov:2009sv}
Mitov A, Sterman G~F and Sung I 2009 {\em Phys. Rev. D\/} {\bf 79} 094015
  (\textit{Preprint} \eprint{0903.3241})

\bibitem{Becher:2009kw}
Becher T and Neubert M 2009 {\em Phys. Rev. D\/} {\bf 79} 125004 [Erratum:
  Phys.Rev.D 80, 109901 (2009)] (\textit{Preprint} \eprint{0904.1021})

\bibitem{Beneke:2009rj}
Beneke M, Falgari P and Schwinn C 2010 {\em Nucl. Phys. B\/} {\bf 828} 69--101
  (\textit{Preprint} \eprint{0907.1443})

\bibitem{Czakon:2009zw}
Czakon M, Mitov A and Sterman G~F 2009 {\em Phys. Rev. D\/} {\bf 80} 074017
  (\textit{Preprint} \eprint{0907.1790})

\bibitem{Ferroglia:2009ii}
Ferroglia A, Neubert M, Pecjak B~D and Yang L~L 2009 {\em JHEP\/} {\bf 11} 062
  (\textit{Preprint} \eprint{0908.3676})

\bibitem{Mitov:2010xw}
Mitov A, Sterman G~F and Sung I 2010 {\em Phys. Rev. D\/} {\bf 82} 034020
  (\textit{Preprint} \eprint{1005.4646})

\bibitem{Gardi:2013saa}
Gardi E 2014 {\em JHEP\/} {\bf 04} 044 (\textit{Preprint} \eprint{1310.5268})

\bibitem{Falcioni:2014pka}
Falcioni G, Gardi E, Harley M, Magnea L and White C~D 2014 {\em JHEP\/} {\bf
  10} 010 (\textit{Preprint} \eprint{1407.3477})

\bibitem{Henn:2013tua}
Henn J~M, Smirnov A~V and Smirnov V~A 2013 {\em JHEP\/} {\bf 07} 128
  (\textit{Preprint} \eprint{1306.2799})

\bibitem{Gardi:2010rn}
Gardi E, Laenen E, Stavenga G and White C~D 2010 {\em JHEP\/} {\bf 11} 155
  (\textit{Preprint} \eprint{1008.0098})

\bibitem{Gardi:2011wa}
Gardi E and White C~D 2011 {\em JHEP\/} {\bf 03} 079 (\textit{Preprint}
  \eprint{1102.0756})

\bibitem{Dukes:2013wa}
Dukes M, Gardi E, Steingrimsson E and White C~D 2013 {\em J. Comb. Theor. A\/}
  {\bf 120} 1012--1037 (\textit{Preprint} \eprint{1301.6576})

\bibitem{Gardi:2013ita}
Gardi E, Smillie J~M and White C~D 2013 {\em JHEP\/} {\bf 06} 088
  (\textit{Preprint} \eprint{1304.7040})

\bibitem{Gardi:2021gzz}
Gardi E, Harley M, Lodin R, Palusa M, Smillie J~M, White C~D and Yeomans S 2021
   (\textit{Preprint} \eprint{2110.01685})

\bibitem{Mitov:2010rp}
Mitov A, Sterman G and Sung I 2010 {\em Phys. Rev. D\/} {\bf 82} 096010
  (\textit{Preprint} \eprint{1008.0099})

\bibitem{Vladimirov:2015fea}
Vladimirov A~A 2015 {\em JHEP\/} {\bf 06} 120 (\textit{Preprint}
  \eprint{1501.03316})

\bibitem{Almelid:2017qju}
Almelid O, Duhr C, Gardi E, McLeod A and White C~D 2017 {\em JHEP\/} {\bf 09}
  073 (\textit{Preprint} \eprint{1706.10162})

\bibitem{Cachazo:2014fwa}
Cachazo F and Strominger A 2014  (\textit{Preprint} \eprint{1404.4091})

\bibitem{Casali:2014xpa}
Casali E 2014 {\em JHEP\/} {\bf 08} 077 (\textit{Preprint} \eprint{1404.5551})

\bibitem{Strominger:2017zoo}
Strominger A 2017  (\textit{Preprint} \eprint{1703.05448})

\bibitem{McLoughlin:2022ljp}
McLoughlin T, Puhm A and Raclariu A~M 2022 {\em J. Phys. A\/} {\bf 55} 443012
  (\textit{Preprint} \eprint{2203.13022})

\bibitem{Kramer:1996iq}
Kramer M, Laenen E and Spira M 1998 {\em Nucl. Phys. B\/} {\bf 511} 523--549
  (\textit{Preprint} \eprint{hep-ph/9611272})

\bibitem{Ball:2013bra}
Ball R~D, Bonvini M, Forte S, Marzani S and Ridolfi G 2013 {\em Nucl. Phys.
  B\/} {\bf 874} 746--772 (\textit{Preprint} \eprint{1303.3590})

\bibitem{Bonvini:2014qga}
Bonvini M, Forte S, Ridolfi G and Rottoli L 2015 {\em JHEP\/} {\bf 01} 046
  (\textit{Preprint} \eprint{1409.0864})

\bibitem{vanBeekveld:2019cks}
van Beekveld M, Beenakker W, Basu R, Laenen E, Misra A and Motylinski P 2019
  {\em Phys. Rev. D\/} {\bf 100} 056009 (\textit{Preprint} \eprint{1905.11771})

\bibitem{vanBeekveld:2021hhv}
van Beekveld M, Laenen E, Sinninghe~Damst\'e J and Vernazza L 2021 {\em JHEP\/}
  {\bf 05} 114 (\textit{Preprint} \eprint{2101.07270})

\bibitem{Low:1958sn}
Low F~E 1958 {\em Phys. Rev.\/} {\bf 110} 974--977

\bibitem{Burnett:1967km}
Burnett T~H and Kroll N~M 1968 {\em Phys. Rev. Lett.\/} {\bf 20} 86

\bibitem{DelDuca:1990gz}
Del~Duca V 1990 {\em Nucl. Phys. B\/} {\bf 345} 369--388

\bibitem{Laenen:2008gt}
Laenen E, Stavenga G and White C~D 2009 {\em JHEP\/} {\bf 03} 054
  (\textit{Preprint} \eprint{0811.2067})

\bibitem{Laenen:2010uz}
Laenen E, Magnea L, Stavenga G and White C~D 2011 {\em JHEP\/} {\bf 01} 141
  (\textit{Preprint} \eprint{1010.1860})

\bibitem{Bonocore:2015esa}
Bonocore D, Laenen E, Magnea L, Melville S, Vernazza L and White C~D 2015 {\em
  JHEP\/} {\bf 06} 008 (\textit{Preprint} \eprint{1503.05156})

\bibitem{Bonocore:2016awd}
Bonocore D, Laenen E, Magnea L, Vernazza L and White C~D 2016 {\em JHEP\/} {\bf
  12} 121 (\textit{Preprint} \eprint{1610.06842})

\bibitem{Bonocore:2020xuj}
Bonocore D 2021 {\em JHEP\/} {\bf 02} 007 (\textit{Preprint}
  \eprint{2009.07863})

\bibitem{Gervais:2017yxv}
Gervais H 2017 {\em Phys. Rev. D\/} {\bf 95} 125009 (\textit{Preprint}
  \eprint{1704.00806})

\bibitem{Gervais:2017zky}
Gervais H 2017 {\em Phys. Rev. D\/} {\bf 96} 065007 (\textit{Preprint}
  \eprint{1706.03453})

\bibitem{Gervais:2017zdb}
Gervais H 2017 {\em {Soft Radiation Theorems at All Loop Order in Quantum Field
  Theory}\/} Ph.D. thesis

\bibitem{Laenen:2020nrt}
Laenen E, Sinninghe~Damst\'e J, Vernazza L, Waalewijn W and Zoppi L 2021 {\em
  Phys. Rev. D\/} {\bf 103} 034022 (\textit{Preprint} \eprint{2008.01736})

\bibitem{Bahjat-Abbas:2019fqa}
Bahjat-Abbas N, Bonocore D, Sinninghe~Damst\'e J, Laenen E, Magnea L, Vernazza
  L and White C~D 2019 {\em JHEP\/} {\bf 11} 002 (\textit{Preprint}
  \eprint{1905.13710})

\bibitem{vanBeekveld:2021mxn}
van Beekveld M, Vernazza L and White C~D 2021  (\textit{Preprint}
  \eprint{2109.09752})

\bibitem{Soar:2009yh}
Soar G, Moch S, Vermaseren J~A~M and Vogt A 2010 {\em Nucl. Phys. B\/} {\bf
  832} 152--227 (\textit{Preprint} \eprint{0912.0369})

\bibitem{Moch:2009hr}
Moch S and Vogt A 2009 {\em JHEP\/} {\bf 11} 099 (\textit{Preprint}
  \eprint{0909.2124})

\bibitem{Moch:2009mu}
Moch S and Vogt A 2009 {\em JHEP\/} {\bf 04} 081 (\textit{Preprint}
  \eprint{0902.2342})

\bibitem{deFlorian:2014vta}
de~Florian D, Mazzitelli J, Moch S and Vogt A 2014 {\em JHEP\/} {\bf 10} 176
  (\textit{Preprint} \eprint{1408.6277})

\bibitem{LoPresti:2014ihe}
Lo~Presti N~A, Almasy A~A and Vogt A 2014 {\em Phys. Lett. B\/} {\bf 737}
  120--123 (\textit{Preprint} \eprint{1407.1553})

\bibitem{DelDuca:2017twk}
Del~Duca V, Laenen E, Magnea L, Vernazza L and White C~D 2017 {\em JHEP\/} {\bf
  11} 057 (\textit{Preprint} \eprint{1706.04018})

\bibitem{vanBeekveld:2019prq}
van Beekveld M, Beenakker W, Laenen E and White C~D 2020 {\em JHEP\/} {\bf 03}
  106 (\textit{Preprint} \eprint{1905.08741})

\bibitem{Bonocore:2014wua}
Bonocore D, Laenen E, Magnea L, Vernazza L and White C~D 2015 {\em Phys. Lett.
  B\/} {\bf 742} 375--382 (\textit{Preprint} \eprint{1410.6406})

\bibitem{Bahjat-Abbas:2018hpv}
Bahjat-Abbas N, Sinninghe~Damst\'e J, Vernazza L and White C~D 2018 {\em
  JHEP\/} {\bf 10} 144 (\textit{Preprint} \eprint{1807.09246})

\bibitem{Ebert:2018lzn}
Ebert M~A, Moult I, Stewart I~W, Tackmann F~J, Vita G and Zhu H~X 2018 {\em
  JHEP\/} {\bf 12} 084 (\textit{Preprint} \eprint{1807.10764})

\bibitem{Boughezal:2018mvf}
Boughezal R, Isgr\`o A and Petriello F 2018 {\em Phys. Rev. D\/} {\bf 97}
  076006 (\textit{Preprint} \eprint{1802.00456})

\bibitem{Boughezal:2019ggi}
Boughezal R, Isgr\`o A and Petriello F 2020 {\em Phys. Rev. D\/} {\bf 101}
  016005 (\textit{Preprint} \eprint{1907.12213})

\bibitem{Kolodrubetz:2016uim}
Kolodrubetz D~W, Moult I and Stewart I~W 2016 {\em JHEP\/} {\bf 05} 139
  (\textit{Preprint} \eprint{1601.02607})

\bibitem{Moult:2016fqy}
Moult I, Rothen L, Stewart I~W, Tackmann F~J and Zhu H~X 2017 {\em Phys. Rev.
  D\/} {\bf 95} 074023 (\textit{Preprint} \eprint{1612.00450})

\bibitem{Feige:2017zci}
Feige I, Kolodrubetz D~W, Moult I and Stewart I~W 2017 {\em JHEP\/} {\bf 11}
  142 (\textit{Preprint} \eprint{1703.03411})

\bibitem{Beneke:2017ztn}
Beneke M, Garny M, Szafron R and Wang J 2018 {\em JHEP\/} {\bf 03} 001
  (\textit{Preprint} \eprint{1712.04416})

\bibitem{Beneke:2018rbh}
Beneke M, Garny M, Szafron R and Wang J 2018 {\em JHEP\/} {\bf 11} 112
  (\textit{Preprint} \eprint{1808.04742})

\bibitem{Bhattacharya:2018vph}
Bhattacharya A, Moult I, Stewart I~W and Vita G 2019 {\em JHEP\/} {\bf 05} 192
  (\textit{Preprint} \eprint{1812.06950})

\bibitem{Beneke:2019kgv}
Beneke M, Garny M, Szafron R and Wang J 2019 {\em JHEP\/} {\bf 09} 101
  (\textit{Preprint} \eprint{1907.05463})

\bibitem{Bodwin:2021epw}
Bodwin G~T, Ee J~H, Lee J and Wang X~P 2021  (\textit{Preprint}
  \eprint{2107.07941})

\bibitem{Moult:2019mog}
Moult I, Stewart I~W and Vita G 2019 {\em JHEP\/} {\bf 11} 153
  (\textit{Preprint} \eprint{1905.07411})

\bibitem{Beneke:2019oqx}
Beneke M, Broggio A, Jaskiewicz S and Vernazza L 2020 {\em JHEP\/} {\bf 07} 078
  (\textit{Preprint} \eprint{1912.01585})

\bibitem{Liu:2019oav}
Liu Z~L and Neubert M 2020 {\em JHEP\/} {\bf 04} 033 (\textit{Preprint}
  \eprint{1912.08818})

\bibitem{Liu:2020tzd}
Liu Z~L, Mecaj B, Neubert M and Wang X 2021 {\em Phys. Rev. D\/} {\bf 104}
  014004 (\textit{Preprint} \eprint{2009.04456})

\bibitem{Boughezal:2016zws}
Boughezal R, Liu X and Petriello F 2017 {\em JHEP\/} {\bf 03} 160
  (\textit{Preprint} \eprint{1612.02911})

\bibitem{Moult:2017rpl}
Moult I, Stewart I~W and Vita G 2017 {\em JHEP\/} {\bf 07} 067
  (\textit{Preprint} \eprint{1703.03408})

\bibitem{Chang:2017atu}
Chang C~H, Stewart I~W and Vita G 2018 {\em JHEP\/} {\bf 04} 041
  (\textit{Preprint} \eprint{1712.04343})

\bibitem{Moult:2018jjd}
Moult I, Stewart I~W, Vita G and Zhu H~X 2018 {\em JHEP\/} {\bf 08} 013
  (\textit{Preprint} \eprint{1804.04665})

\bibitem{Beneke:2018gvs}
Beneke M, Broggio A, Garny M, Jaskiewicz S, Szafron R, Vernazza L and Wang J
  2019 {\em JHEP\/} {\bf 03} 043 (\textit{Preprint} \eprint{1809.10631})

\bibitem{Ebert:2018gsn}
Ebert M~A, Moult I, Stewart I~W, Tackmann F~J, Vita G and Zhu H~X 2019 {\em
  JHEP\/} {\bf 04} 123 (\textit{Preprint} \eprint{1812.08189})

\bibitem{Beneke:2019mua}
Beneke M, Garny M, Jaskiewicz S, Szafron R, Vernazza L and Wang J 2020 {\em
  JHEP\/} {\bf 01} 094 (\textit{Preprint} \eprint{1910.12685})

\bibitem{Moult:2019uhz}
Moult I, Stewart I~W, Vita G and Zhu H~X 2020 {\em JHEP\/} {\bf 05} 089
  (\textit{Preprint} \eprint{1910.14038})

\bibitem{Liu:2020ydl}
Liu Z~L and Neubert M 2020 {\em JHEP\/} {\bf 06} 060 (\textit{Preprint}
  \eprint{2003.03393})

\bibitem{Liu:2020eqe}
Liu Z~L, Mecaj B, Neubert M, Wang X and Fleming S 2020 {\em JHEP\/} {\bf 07}
  104 (\textit{Preprint} \eprint{2005.03013})

\bibitem{Wang:2019mym}
Wang J 2019  (\textit{Preprint} \eprint{1912.09920})

\bibitem{Beneke:2020ibj}
Beneke M, Garny M, Jaskiewicz S, Szafron R, Vernazza L and Wang J 2020 {\em
  JHEP\/} {\bf 10} 196 (\textit{Preprint} \eprint{2008.04943})

\bibitem{White:2006yh}
White C~D and Thorne R~S 2007 {\em Phys. Rev. D\/} {\bf 75} 034005
  (\textit{Preprint} \eprint{hep-ph/0611204})

\bibitem{Ball:2017otu}
Ball R~D, Bertone V, Bonvini M, Marzani S, Rojo J and Rottoli L 2018 {\em Eur.
  Phys. J. C\/} {\bf 78} 321 (\textit{Preprint} \eprint{1710.05935})

\bibitem{Andersen:2020yax}
Andersen J~R, Black J~A, Brooks H~M, Byrne E~P, Maier A and Smillie J~M 2021
  {\em JHEP\/} {\bf 04} 105 (\textit{Preprint} \eprint{2012.10310})

\bibitem{Andersen:2019yzo}
Andersen J~R, Hapola T, Heil M, Maier A and Smillie J 2019  (\textit{Preprint}
  \eprint{1902.08430})

\bibitem{Andersen:2018kjg}
Andersen J~R, Cockburn J~D, Heil M, Maier A and Smillie J~M 2019 {\em JHEP\/}
  {\bf 04} 127 (\textit{Preprint} \eprint{1812.08072})

\bibitem{Andersen:2018tnm}
Andersen J~R, Hapola T, Heil M, Maier A and Smillie J~M 2018 {\em JHEP\/} {\bf
  08} 090 (\textit{Preprint} \eprint{1805.04446})

\bibitem{Andersen:2017kfc}
Andersen J~R, Hapola T, Maier A and Smillie J~M 2017 {\em JHEP\/} {\bf 09} 065
  (\textit{Preprint} \eprint{1706.01002})

\bibitem{Andersen:2016vkp}
Andersen J~R, Medley J~J and Smillie J~M 2016 {\em JHEP\/} {\bf 05} 136
  (\textit{Preprint} \eprint{1603.05460})

\bibitem{Andersen:2012gk}
Andersen J~R, Hapola T and Smillie J~M 2012 {\em JHEP\/} {\bf 09} 047
  (\textit{Preprint} \eprint{1206.6763})

\bibitem{Andersen:2011hs}
Andersen J~R and Smillie J~M 2011 {\em JHEP\/} {\bf 06} 010 (\textit{Preprint}
  \eprint{1101.5394})

\bibitem{Andersen:2009nu}
Andersen J~R and Smillie J~M 2010 {\em JHEP\/} {\bf 01} 039 (\textit{Preprint}
  \eprint{0908.2786})

\bibitem{Eden:1966dnq}
Eden R~J, Landshoff P~V, Olive D~I and Polkinghorne J~C 1966 {\em {The analytic
  S-matrix}\/} (Cambridge: Cambridge Univ. Press)

\bibitem{Collins:1977jy}
Collins P~D~B 2009 {\em {An Introduction to Regge Theory and High-Energy
  Physics}\/} Cambridge Monographs on Mathematical Physics (Cambridge, UK:
  Cambridge Univ. Press) ISBN 978-0-521-11035-8

\bibitem{White:2019ggo}
White C~D 2020 {\em SciPost Phys. Lect. Notes\/} {\bf 13} 1 (\textit{Preprint}
  \eprint{1909.05177})

\bibitem{ATLAS:2017qey}
Aaboud M {\em et~al.\/} (ATLAS) 2017 {\em JHEP\/} {\bf 10} 132
  (\textit{Preprint} \eprint{1708.02810})

\bibitem{Ellis:1996mzs}
Ellis R~K, Stirling W~J and Webber B~R 2011 {\em {QCD and collider physics}\/}
  vol~8 (Cambridge University Press) ISBN 978-0-511-82328-2, 978-0-521-54589-1

\bibitem{Caravaglios:1998yr}
Caravaglios F, Mangano M~L, Moretti M and Pittau R 1999 {\em Nucl. Phys. B\/}
  {\bf 539} 215--232 (\textit{Preprint} \eprint{hep-ph/9807570})

\bibitem{Catani:2001cc}
Catani S, Krauss F, Kuhn R and Webber B~R 2001 {\em JHEP\/} {\bf 11} 063
  (\textit{Preprint} \eprint{hep-ph/0109231})

\bibitem{Frixione:2002ik}
Frixione S and Webber B~R 2002 {\em JHEP\/} {\bf 06} 029 (\textit{Preprint}
  \eprint{hep-ph/0204244})

\bibitem{Frixione:2003ei}
Frixione S, Nason P and Webber B~R 2003 {\em JHEP\/} {\bf 08} 007
  (\textit{Preprint} \eprint{hep-ph/0305252})

\bibitem{Re:2021vzu}
Re E 2021 {Recent progress on high order calculations and matching to parton
  showers} {\em {9th Large Hadron Collider Physics Conference}\/}
  (\textit{Preprint} \eprint{2110.02183})

\bibitem{Sjostrand:2019zhc}
Sj\"ostrand T 2020 {\em Comput. Phys. Commun.\/} {\bf 246} 106910
  (\textit{Preprint} \eprint{1907.09874})

\bibitem{Salam:2010nqg}
Salam G~P 2010 {\em Eur. Phys. J. C\/} {\bf 67} 637--686 (\textit{Preprint}
  \eprint{0906.1833})

\bibitem{HEPData}
{HEPData}: Repository for publication-related high-energy physics data
  \url{https://www.hepdata.net} accessed: 2022-06-27

\end{thebibliography}

\end{document}